# Measurement of Spectrum Averaged Cross Sections in LR-0 Benchmark Reference Neutron Field

Michal Košťál[1], Evžen Losa[1], Stanislav Simakov[2], Tomáš Czakoj[1], Martin Schulc[1], Jan Šimon[1], Vojtěch Rypar[1], Martin Mareček[1], Jan Uhlíř[1], Alena Krechlerová[1], Tomáš Peltan[1], Radek Pošvař[3], Zdeněk Matěj[4], David Bernard[5], Andrej Trkov[6], Roberto Capote[7]

[1] Research Center Rez, 250 68 Husinec-Rez 130, Czech Republic
[2] Institute for Neutron Physics and Reactor Technology, Karlsruhe Institute of Technology, Hermann-von-Helmholtz-Platz 1, D-76344 Eggenstein-Leopoldshafen, Germany
[3] UJV Rez, 250 68 Husinec-Rez 130, Czech Republic
[4] Masaryk University, Botanická 15, Brno 612 00, Czech Republic
[5] CEA/DES/IRESNE/DER/SPRC, Laboratoire d'Etudes de Physique (LEPh), Cadarache, 13108, Saint Paul-lez-Durance, France
[6] Jožef Stefan Institute, Jamova cesta 39, 1000 Ljubljana, Slovenia
[7] NAPC-Nuclear Data Section, International Atomic Energy Agency, A-1400 Wien, Austria

Email: Michal.Kostal@cvrez.cz
Telephone: +420266172655

Abstract

The measured and evaluated excitation functions are fundamental quantities that affect the accuracy of all calculations in nuclear applications. Some cross sections, such as $^{14}$N(n,p)$^{14}$C, have added value for special applications, as these reactions may be responsible for possible contamination in industrial processes such as spent fuel reprocessing. For the validation of the evaluated cross sections, we can rely on the comparison of the calculated spectrum averaged cross sections (SACS) for the given neutron spectrum with the measured SACS value. The benchmark reference neutron field has been identified, characterized, and well validated in the LR-0 special core. A very large set of SACS measurements in the LR0 reference field is measured with low uncertainty, making it an excellent set for deconvolution codes' validation. The impact of the gamma-induced reaction on the production yield of neutron-induced reactions was investigated for most of the benchmarked reactions. Gamma competition was found to contribute at most 1% for the $^{197}$Au(n,2n) reaction, while being substantially lower for other target isotopes and neutron-induced activation reactions.

## 1 Introduction

Historically, several irradiation experiments were carried out in the benchmark neutron field in the special core of the LR-0 reactor. Its neutron field was identified to be well-characterized, stable in time, and reproducible; therefore, it could be declared as a reference neutron field (see Kostal et al., 2020). Reference neutron fields are especially useful for integral experiments aimed at validations of neutron cross-section evaluations and the performed experiments served this purpose.

The output of the irradiation experiment is activated material, from which the reaction rate and spectrum averaged cross section (SACS) can be evaluated if the neutron spectrum shape is known. Because each material and each reaction can have different sensitivity to different neutron energy, the compiled set of measured activation reactions can serve for neutron spectrum characterization and for example, for testing of spectrum deconvolution codes.

Current study summarizes past experiments and introduces a new one covering a very wide neutron energy range, as shown further in chapter. These experiments were performed under slightly different circumstances. So, they have minor differences, such as slight changes in the neutron spectrum. In the next sections, it is shown that all the experiments performed can be normalized by neutron flux above 10 MeV, so that they are y comparable to each other.

Previous studies (Kostal et al 2017b, Kostal et al., 2020) refer to experimental work carried out to validate different threshold reactions evaluations possible for fast neutron flux monitoring. Current work introduces the methodology for renormalization of the older results, which then do not need to be modified by different corrections such as flux loss or spectral shift correction. It was found that sometimes the (γ,n) reaction channel can significantly contribute to the evaluation of (n,2n) channel because the reaction product is identical. The influence of the gamma flux on the evaluation of neutron reactions is evaluated in chapter 4.5.

It was shown previously that the neutron spectrum in the benchmark reference neutron field is undistinguishable from $^{235}$U PFNS in region above 6 MeV (Kostal et al., 2018). Current paper extends that investigation by examining the impact of individual core components on this similarity, as well as the effect of deviations from the $^{235}U(n_{th},f)$ PFNS on Spectrum Average Cross Sections (SACS) for reactions with thresholds below 6 MeV.

One of the main goals of this study was the validation of the $^{14}N(n,p)^{14}C$ reaction (see chapter 4.1), which is one of the important reactions because neutron activated nitrogen impurities are responsible for $^{14}$C contamination of nuclear fuel, which then has implications for spent fuel treatment in reprocessing plants Plumbet al 1984. This reaction needs to be considered in astrophysics because it plays a crucial role as a neutron poison for S-process nucleosynthesis Brehm et al 1988. Moreover, accurate estimation of this reaction is vital for correct dose estimation during boron neutron capture therapy Skwierawska et al 2022.

Due to possible obstacles with the free form of nitrogen, a suitable nitrogen-rich compound had to be chosen for irradiation experiment in the LR-0 reactor. A suitable candidate for this purpose could be alanine (2-Aminopropionic acid), which also serves as a relevant radiation damage monitor of the living tissues (Grasso et al 2023). A large set of foils was irradiated in several LR-0 cycles together with alanine as the nitrogen source. Irradiation was carried out in three extensive experiments that support previous measurement of a large set of spectrum averaged cross sections useful for validation dosimetry cross sections.

# 2 LR-0 reactor

The LR-0 research reactor is a light-water, zero-power pool-type reactor operated by the Research Centre Rez. It uses VVER-1000 type fuel. One assembly consists of 312 fuel pins in a triangular lattice with a pitch of 12.75 mm. Fuel pellets, in the form of $UO_2$ with a density of ≈ 10.33 g/cm$^3$, are placed inside zirconium-alloy cladding tubes with an inner diameter of 7.73 mm and an outer diameter of 9.15 mm, respectively. A hole with a diameter of 1.4 mm is drilled in the center of each fuel pellet with an outer diameter of 7.53 mm. Each fuel assembly contains 18 absorber cluster tubes made of stainless steel and a central zirconium-alloy tube. The fuel grid has a pitch of 23.6 cm. Reactor criticality is driven by the water level; however, water level changes during irradiation are very small, not higher than 0.5 mm.

The special core used in experiments consists of six fuel assemblies with an enrichment of 3.3%, arranged around the dry hexagonal channel with geometrical cross section nearly identical to that of the fuel (a hexagon with an apothem of 11.7 cm). The radial center and axial level (see $H_{ref}$ in Table 2), where the used samples are irradiated (see Figure 1, left), is placed within this dry hexagonal channel.

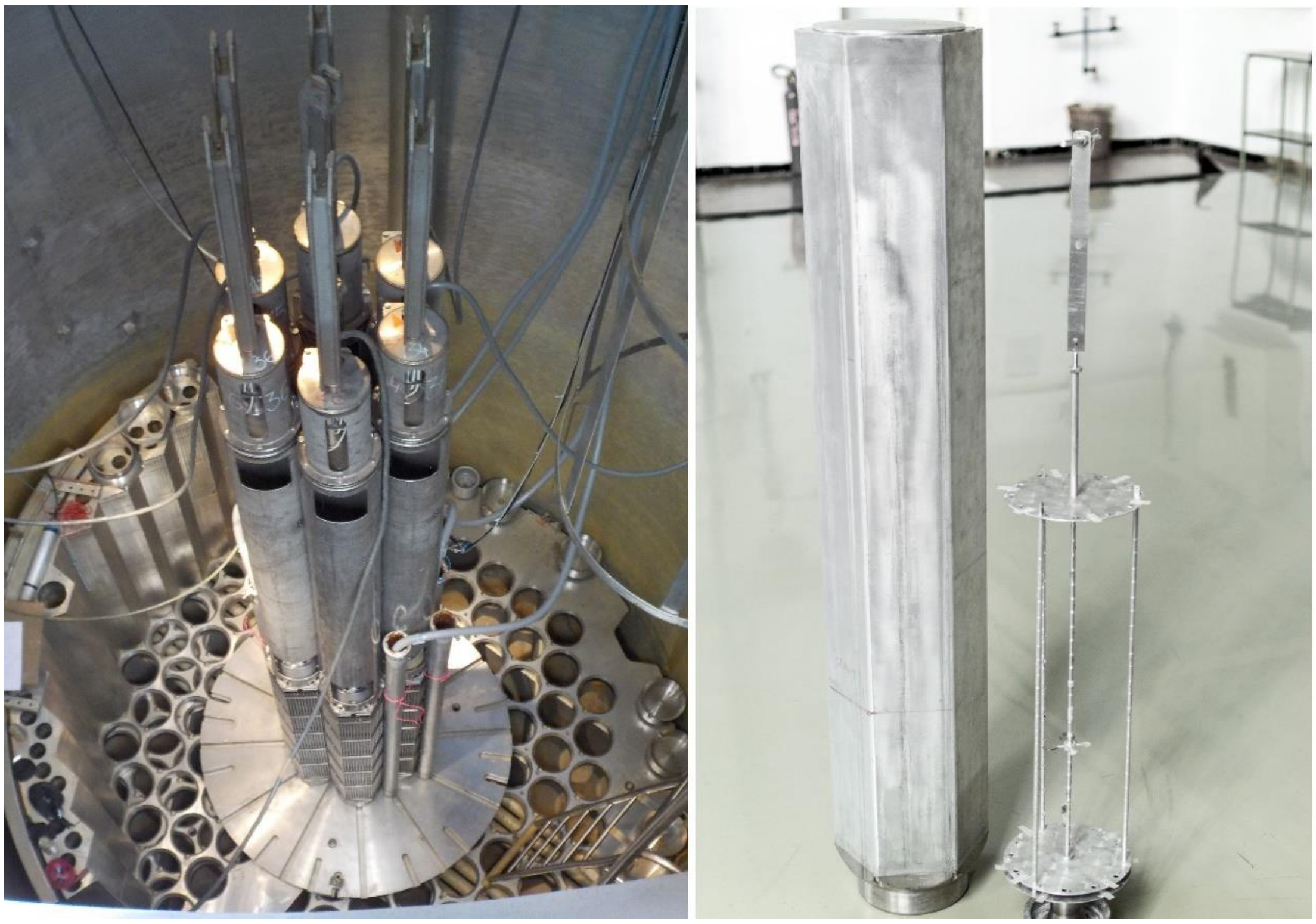

Figure 1.: View on the LR-0 reactor special core (left) with aluminum dry hexagonal channel placed in core center where benchmark reference neutron field was identified (right) (Trkov et al., 2020)

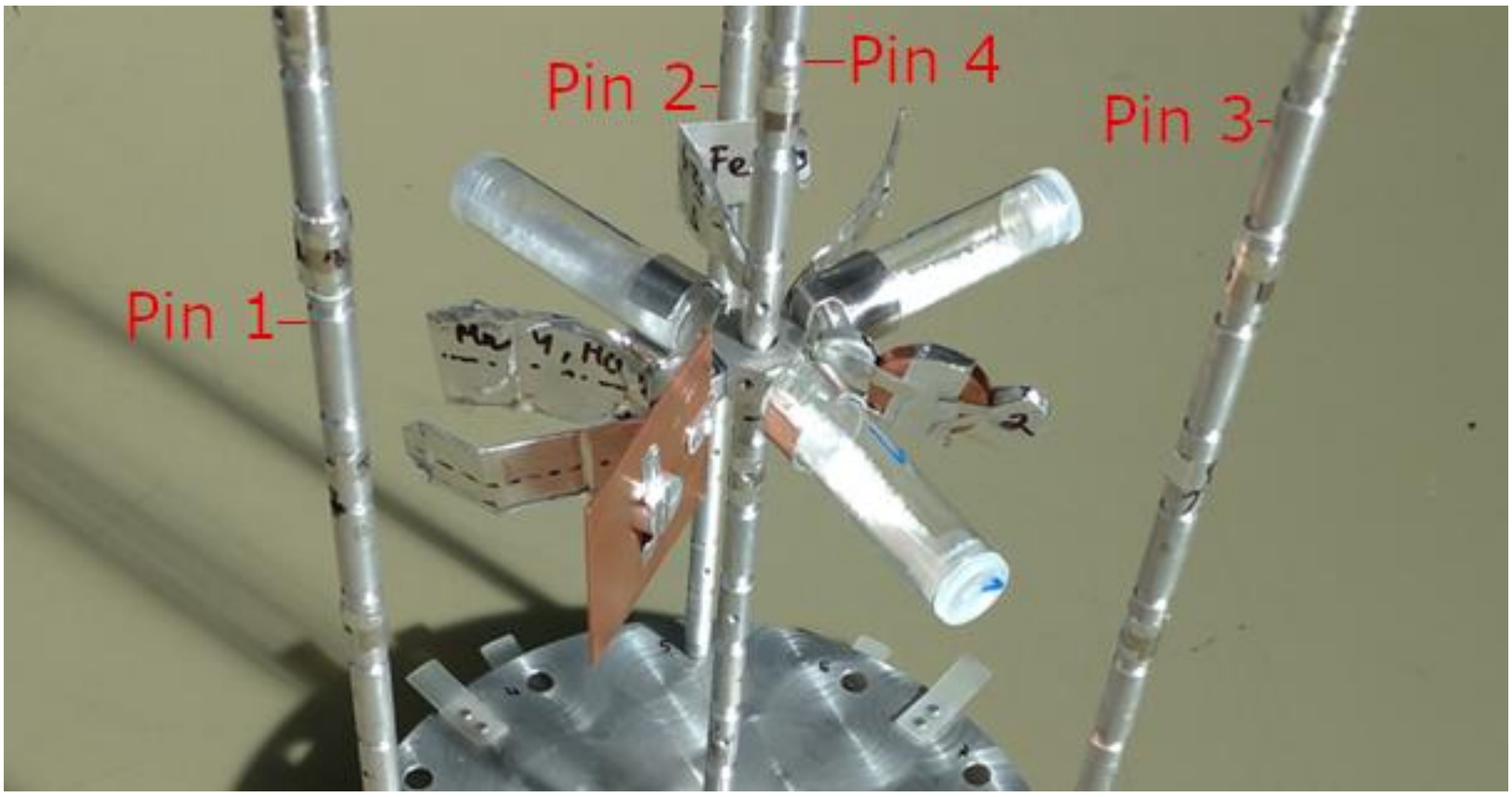

Figure 2.: Detailed view on the target formed by the alanine samples and large set of foils in 2020 experimental set

# 3 Experimental and calculation methods

## 3.1 Irradiation setup

Three new experiments were carried out with small foils and reevaluation was performed for six older experiments with large volume samples. These were renormalized to form a single set of large-scale spectrum-averaged cross sections (SACS). Large samples were used together with activation foil monitors stuck to the sample.

Alanine was selected as a nitrogen-containing agent for its chemical stability both during scintillation measurements of irradiated sample and throughout the irradiation process. There are other materials with higher nitrogen abundance , such as hydrazine or urea, but their use was excluded due to their highly oxidizing properties when mixed with used scintillation cocktail. To prevent the leakage of $^{14}C$ during irradiation, the alanine was encapsulated in a glass vial. An extensive set of foils in stable metallic form was irradiated together with alanine, a detailed view is shown Figure 2.

It was shown experimentally that the neutron spectrum in the used neutron field is identical with $^{235}U(n_{th}, f)$ prompt fission neutron spectrum (PFNS) in the 6–14 MeV region (Kostal et al., 2018), while calculation confirmed this in the 6–20 MeV region (Kostal et al. 2022b). However, due to the design of the driver reference core and variations in water level or even vertical positioning of the cylindrical volume containing samples, shifts in neutron spectrum have been observed among three consecutive irradiation sets. Figure 3 illustrates the comparison of neutron spectra from experiments carried out between 2020 and 2022, which featured different experiment layouts. Variations included the positioning of foils within the holder, placement of instrumentation tubes (outside the core), and slight differences in moderator level. The spectra were calculated using the MCNP code and normalized in the region above 1 eV. Naturally, according to the previously verified identity of the spectra with $^{235}U(n_{th},f)$ PFNS, all studied spectra in the region above 6 MeV exhibit almost identical characteristics.

Some variations of the spectra occur in the region below 0.2 eV, with deviations less than 2 %; however in several groups in the resonance region around 300 eV, these deviations can reach up to 10 % (refer to Figure 3). Indeed, for reactions with a threshold above 0.1 MeV, where the spectrum was well-validated (Kostal et al., 2015, Kostal et al., 2017, Kostal et al., 2017a), the computationally evaluated spectral shift effect does not exceed calculational uncertainties, typically on the order of 0.1%. This leads to the conclusion that the evaluations of spectrum averaged cross sections of threshold reactions in various arrangements can be combined, only in the case of capture reactions they must be corrected for the spectral shift effect.

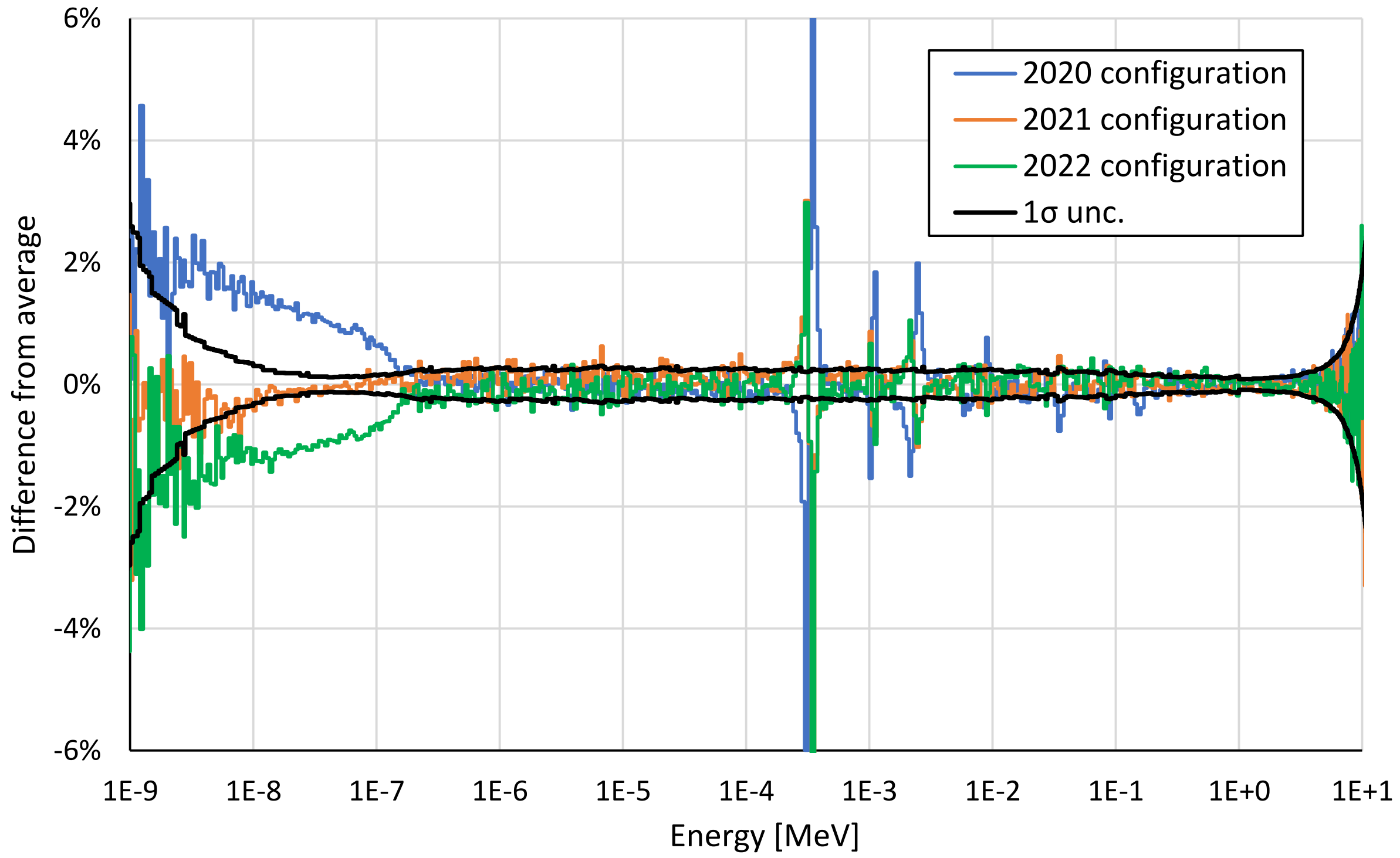

Figure 3.: Comparison of MCNP calculated spectra in various special core arrangements together with related uncertainty (1 σ uncertainty level).

Because of slight variations in reactivity during operations, the reactor power is not constant but can be described as piecewise continuous. Consequently, the power evolution during the experiment was considered using the ratio of the saturated activity in the selected power density and is defined in Equation (1).

$$\frac{A(\overline{P})}{A_{\mathrm{Sat}}(\overline{P})} = \sum_i \frac{P_i}{\overline{P}} \times \left(1 - e^{-\lambda \cdot T_i}\right) \times e^{-\lambda \cdot T_i^{\mathrm{End}}} \qquad (1)$$

Here A/A$_{sat}$ is the ratio of the activity to saturated activity, $\frac{P_i}{\overline{P}}$ is the relative power in i-th interval of the irradiation period, $\overline{P}$ is the average power level in the first day of the irradiation experiment; $\lambda_i$ is the decay constant of the radioisotope considered, $T_i$is the irradiation time in i-th interval of the irradiation period and $T_i^{\mathrm{End}}$ is the time from the end of the i-th irradiation interval to the end of irradiation period.

### 3.2 Gamma spectrometry

The experimental neutron activation reaction rates, described by Equation (2), were determined based on the measured induced activity of the dosimetry foils. This quantity was derived from Net Peaks Area measured employing semiconductor HPGe gamma spectrometry.

The measurement was carried out using a well-characterized HPGe spectrometric system in the Research Centre Rez (see Kostal et al., 2018b). To minimize geometrical uncertainty in the measurements of irradiated samples, thin foils, namely Mg, Fe, $^{54}$Fe, Mo and Ni monitors, were fixed in the plastic EG-3 holder (special plastic holder, for more details, see Fig. 10 in Matej et al 2022). These foils within the holder were placed atop a coaxial HPGe detector oriented vertically to maximize the measured response. The masses of samples were optimized based on the expected activity at the end of irradiation ranging from approximately 0.2 grams up to a few grams.

In older experiments focused on measurement of high threshold cross sections, large volume samples were used. For instance, in the case of sodium, where cross section is very low and half-life of the measured product is large, the sample consisted of of 644 g of NaF powder. To enhance measuring efficiency, the NaF powder was placed into a Marinelli beaker (Kostal et al 2019). The detection efficiencies were determined through calculations, made possible by thorough characterization of the detection system, as discussed below.

The detector energy calibration was conducted prior to the experiment using standard point sources $^{60}Co$, $^{88}Y$, $^{133}Ba$, $^{137}Cs$, $^{152}Eu$, and $^{241}Am$. An energy uncertainty of less than 1.0 keV was achieved across the energy range used. The efficiency curve and Coincidence Summing Factors (Tomarchio et al., 2009 ) were calculated using the MCNP6 code and validated mathematical model of HPGe. The validation was performed for all used geometries, including the point source on the detector cap, the point source positioned 10 cm above the cap, and the Marinelli beaker representing a large volume source. For the point source on the cap, the discrepancy between calculation and experiment in the relevant gamma energy region was below 1.8 %. In other cases, such as the point source positioned 10 cm above the cap and the Marinelli beaker, the discrepancy was below 1 %.

$$q(\overline{P}) = \left(\frac{A(\overline{P})}{A_{\mathrm{Sat}}(\overline{P})}\right)^{-1} \times \frac{NPA}{T_{Live}} \times \frac{1}{\varepsilon \times \eta \times N} \times \frac{\lambda \times T_{Real}}{(1 - e^{-\lambda \cdot T_{Real}})} \times \frac{1}{e^{-\lambda \cdot \Delta T}} \times \frac{1}{CSCF} \qquad (2)$$

Where:

$q(\overline{P})$; is the reaction rate of activation during power density at the power level P$\overline{P}$;

$\overline{P}$ average power level in the first day of the irradiation experiment;

$NPA$; is the measured Net Peak Area$T_{Live}$; is time of measurement by HPGe;

$T_{Real}$; is time of measurement by HPGe, corrected to detector dead time;

$T_{live}$ is the live time in the HPGe measurement;

$\Delta T$; is the time between the end of irradiation and the start of HPGe measurement;

$\lambda$; is decay constant of studied isotope;

$\varepsilon$; is the gamma branching ratio;

$\eta$; is the detector efficiency (the result of MCNP6 calculation);

$N$; is the number of target isotope nuclei;

$CSCF$; is the coincidence summing correction factor.

### 3.3 Scintillation counting in $^{14}C$ measurement

The activity of $^{14}C$ was determined using liquid scintillation counting. A small amount of irradiated alanine sample 0.5 – 1 g was dissolved in 10 ml of low background water and mixed with 10 ml of scintillation cocktail GOLDSTAR LT[2] in a plastic scintillation vial. Background sample and two standards of $^{14}C$ reference material (RM) were prepared in the same way. This set of samples were measured in the Central analytical laboratory (ÚJV Řež a. s.) using low-background liquid scintillation spectrometer Quantulus GCT 6220. The activity of $^{14}C$ with combined uncertainty (0.1265 ± 6.23 %) Bq/g in irradiated alanine was determined in the optimized spectra region.

The possibility of mismatch caused by differences in vials preparation between standards and sample was eliminated adding known small amount of $^{14}C$ RM to each sample vial. The activity of $^{14}C$ with combined uncertainty (0.1206 ± 6.19 %) Bq/g was determined this way by the method of internal standard. The difference between these determinations is negligible according to the uncertainties, which proves the good accuracy and repeatability of this method.

### 3.4 Scintillation spectrometry in gamma spectra measurement

The gamma spectrum in region 1 MeV to 13 MeV was measured using stilbene scintillation detector 45 × 45 mm (diameter × height), which was connected to the NGA-01 spectrometer. The experiment was realized in 3.6% enriched core due to very low background in this fuel type in comparison with previous referenced experiments using core with 3.3 % fuel.

Details about the spectrometer are available in Czakoj et al., 2022. Neutron gamma discrimination was performed using pulse shape discrimination (PSD). The gamma spectrum was then obtained via deconvolution using the Maximum Likelihood Estimation (see Cvachovec, J., Cvachovec, F., 2008).

### 3.5 Calculation methods

The simulations of neutron and photon transport from the core to the irradiated target were performed in criticality calculations using the MCNP6 and TRIPOLI codes. The gamma transport from the activated target to HPGe was calculated using the MCNP6.

#### 3.5.1 The MCNP simulations

The MCNP simulations utilized MCNP6.2 (Werner et al., 2018) code coupled with the ENDF/B-VIII.0 (Brown et al., 2018) and ENDF/B-VII.1 (Chadwick et al. 2011) nuclear data libraries for the neutron and gamma transport during the critical cycles and also dosimetry cross sections from IRDFF-II (Trkov et al., 2020) library, JEFF-3.3. (Plompen et al 2020) or JENDL-5 library (Iwamoto et al 2023) for the activation calculations of foils or targets. The model of the core was assembled according to the OECD NEA benchmark report (Kostal et al., 2017a).

The critical parameter at the irradiation phase was the neutron moderator level, which was adjusted in each experiment due to the different sample dimensions and positions. The critical levels related to the different experiments are summarized in Table 2 for experiments with small foils placed on special Al holder and in Table 5 for large samples.

The reaction rates or neutron spectra used for reaction rate calculation were scored at the exact position of the materials. To evaluate the flux loss and spectrum shift, cases with and without material were modeled as well. In the case of neutron spectrum calculations, the SAND-II 640 group structure was used for further reaction rate calculations.

Usually, $2\times10^4$ cycles with $10^6$ neutrons were sufficient to reach sufficient accuracy of reactions with thresholds above 10 MeV. In the case of gamma spectrum determination, longer calculations in coupled neutron-photon transport in the critical mode were used to reach satisfactory statistics in the high energy photon region of interest.

#### 3.5.2 TRIPOLI simulations

The TRIPOLI-4® (Brun et al. 2015) input desk was converted on the bases of the previously described MCNP input desk. In that sense, a perfect reliability is reached for this surface and native description of LR-0 reactor core with TRIPOLI-4®. The nuclear data library used for neutron transport are JEFF-3.1.1 (OECD 2009) and JEFF-4T2 for pointwise cross-sections for resonant ranges and probability tables describing unresolved resonance ranges. The atomic data for photon transport is EPDL-92. HPC calculations are done with more than 200 000 batches of 20 000 neutrons on 512 processors during 24h reaching about billions of simulated neutrons and gammas. For scoring, natural JEFF cross sections are used instead of multigroup response functions for the IRDFF-II library.

## 4 Results

### 4.1 Evaluation of $^{14}N(n,p)^{14}C$

The spectrum averaged cross section of the $^{14}N(n,p)^{14}C$ reaction was compared with the theoretical prediction using a set of nuclear data libraries (see Table 1). It is worth noting that the values obtained from the calculations using MCNP are in good agreement with the experiment. In case of JENDL-5 the difference is below 1%, in the case of ENDF/B-VIII.0 and JEFF-3.3 – within up to 5 %. The experimental uncertainty covers the uncertainty in measurement technique, and uncertainty related to used glass vial composition used in irradiation.

Table 1.: Experimental and calculated SACS for the $^{14}N(n,p)^{14}C$ reaction averaged in the LR-0 spectrum

| | Mean [b] | Unc. | C/E-1 |
|---|---|---|---|
| Experiment | 0.3067 | 10.4% | - |
| ENDF/B-VIII.0 | 0.2920 | 0.10% | -4.8% |
| JEFF-3.3 | 0.2920 | 0.10% | -4.8% |
| JENDL-5 | 0.3060 | 0.10% | -0.2% |

## 4.2 Neutron and gamma fluxes

The calculated neutron and gamma fluxes normalized per 1 source neutron at the reference position are plotted in Figure 4. It is important to mention that the neutron spectra calculated with ENDF/B-VIII.0 and previous evaluation with ENDF/B-VII.1 (Chadwick et al. 2011) are very close in the region below 10 MeV, but differ in the higher energy region, where the ENDF/B-VIII.0 spectrum is higher than ENDF/B-VII.1 or JEFF-3.1.1. The total gamma flux is higher than the neutron flux in the region above 3 MeV; but below this area it is the opposite.

The new measurement of the gamma spectrum in a 3.6 % enriched core was carried out. This core was chosen due to very low background activity, which allows such measurement. It was also shown that the shapes of both neutron and gamma spectra in reference (3.3 %) and 3.6 % core are very similar (Kostal et al., 2017). The comparison between the measured and calculated gamma spectra is plotted in Figure 5, and a satisfactory agreement was reached. This leads to the conclusion that the calculated gamma spectrum is reliable and could be used as input for the estimation of the competition between the gamma and neutron induced reactions, see Section 4.5.

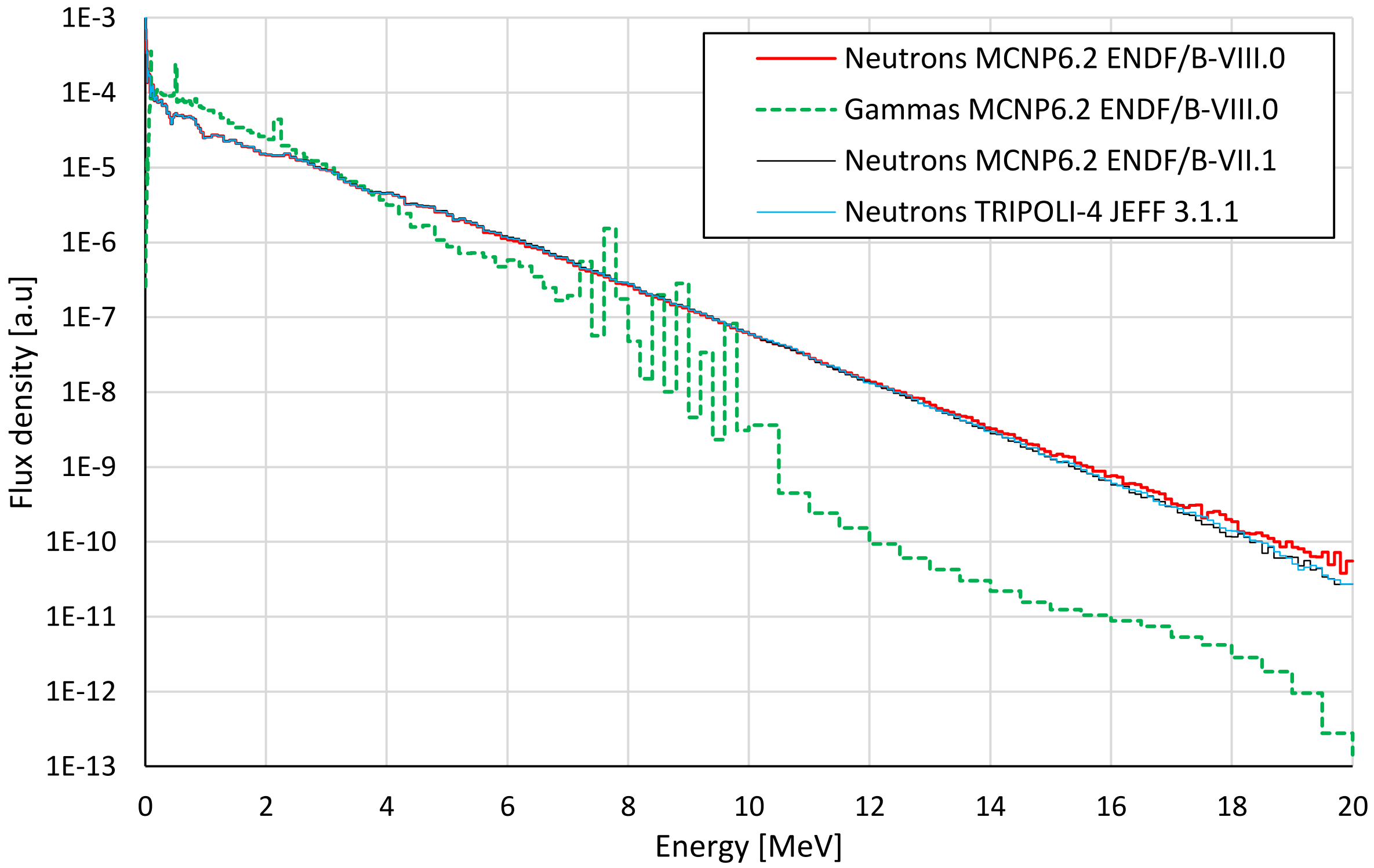

Figure 4.: Calculated neutron and gamma fluxes in MCNP 6.2 and TRIPOLI

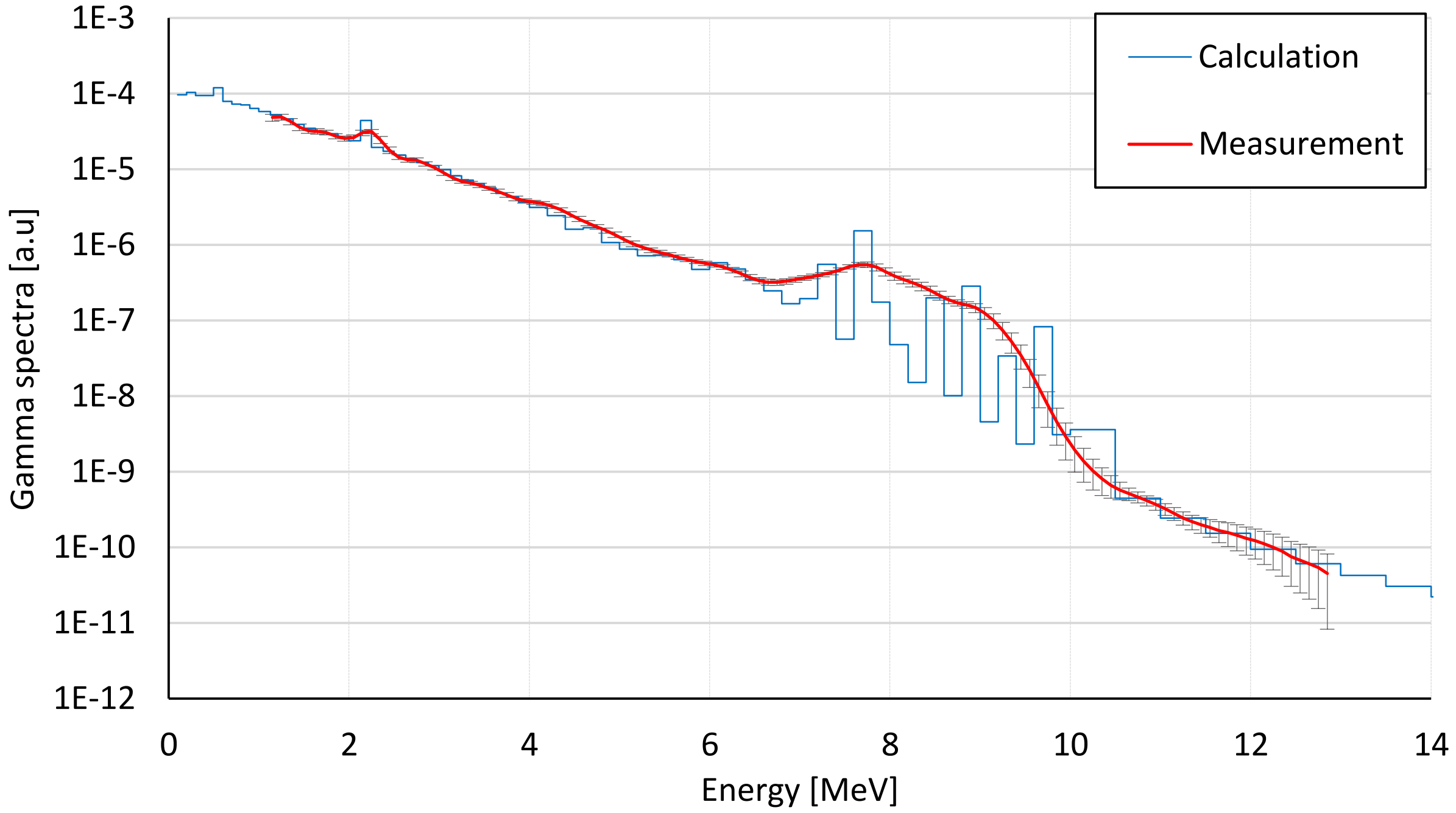

Figure 5.: Validation of gamma flux below 13 MeV in the 3.6 % enriched core calculated in ENDF/B-VIII.0

For a more detailed analysis of the high energy part of the LR-0 neutron spectrum, the ratio of the calculated MCNP spectrum to the $^{235}U(n_{th},f)$ prompt fission neutron spectrum from ENDF/B-VIII.0 is shown in Figure 6. Such comparison indicates differences in the lower neutron energy intervals and rather similar energy shapes above ≈ 6 MeV. Practically identical energy trends is observed for the LR-0 and $^{235}U(n_{th},f)$ PFN spectra from 6 MeV up to 20 MeV, but with local and statistically significant fluctuations.

For better visualization and therefore for the reduction of the Monte-Carlo statistical uncertainties, the LR-0/$^{235}U(n_{th},f)$ spectral ratio above 10 MeV was linearly smoothed over 9 neutron energy bins (or ≈ 1 MeV interval). This allowed us to better see that the shapes of both compared neutron spectra are practically identical up to 16 – 19 MeV.

It is worth noting that the difference between the shapes of the VR-1 reactor (another zero power light water moderated reactor, more details are in Kostal et al 2021) neutron spectrum and $^{235}U(n_{th},f)$ fission neutron spectrum was found to be significantly larger, which was explained by the greater thickness of cooling water between the uranium fuel rods and foil irradiation point (Kostal et al. 2022b).

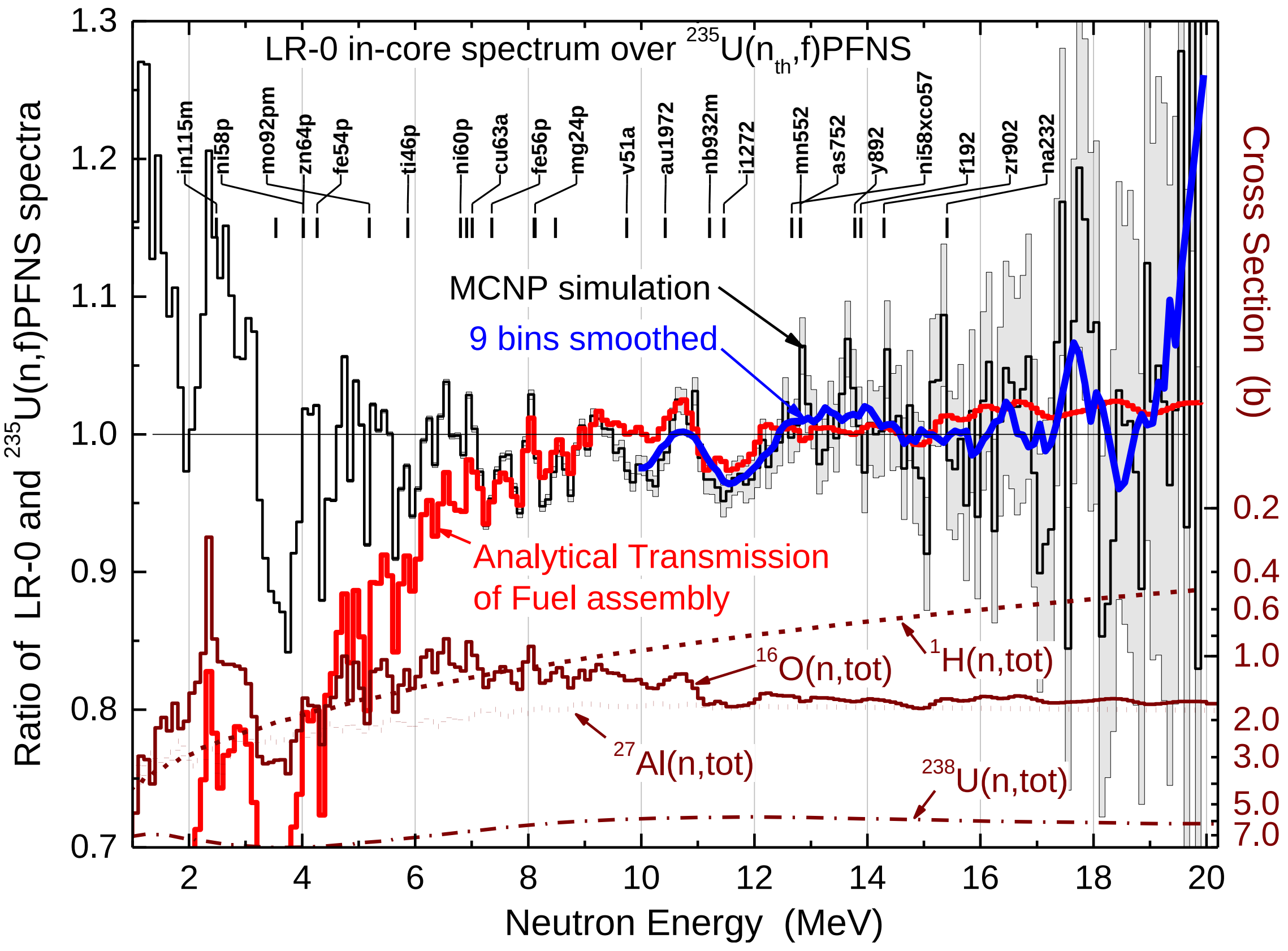

Figure 6.: Left hand axis: Black - Ratio of the MCNP calculated LR-0 neutron spectrum over $^{235}$U(n_th,f) PFNS, Right hand axis: Red - Analytically calculated neutron transmission through the fuel assembly cell and aluminum tube of the dry irradiation hexagonal channel, Grey - total neutron cross sections for neutron interaction with $^{1}$H, $^{16}$O, $^{27}$Al and $^{238}$U (note the downward direction). Abscissa: The median n-SACS energies E50% for the studied reactions are indicated by vertical bars with the reaction shorthand labels adopted in IRDFF (Trkov et al., 2020).

To understand the origin of the same energy trends of the LR-0 and $^{235}$U($n_{th}$,f) spectra and the local oscillations in the case of LR-0, the neutron transmission through the fuel assembly element (3.3 wt.% enriched $UO_2$ rods in Zr cladding surrounded by water) and the Al tube of the dry hexagonal channel was analytically calculated. The neutron total cross sections for the involved materials were taken from ENDF/B-VIII.0. They were processed into 725 energy groups by the PREPRO 2023 code (Cullen 2023) for the present calculations. The exponential attenuation was then computed with the effective thickness of 3.1 mm for $UO_2$ pellets, 0.8 mm for Zr cladding, ≈ 7 mm for water, and 6 mm for the total aluminum layers.

Figure 6 shows comparison of the analytically calculated neutron transmission and the ratio of LR-0 spectrum to $^{235}$U($n_{th}$,f). As can be seen, the analytical approach reproduces quite well the behavior obtained by exact Monte-Carlo simulations. The very similar overall energy trends of the LR-0 and pure $^{235}$U($n_{th}$,f) neutron spectra can now be attributed to the energy invariance of the neutron total cross sections for uranium, aluminum, and oxygen above ≈ (8 – 10) MeV. It can also be seen that the fine structure in the LR-0 spectrum below ≈ 13 MeV is determined by the energy fluctuating of the total cross section for $^{16}$O, which is the element of the reactor fuel and cooling water.

An important consequence of the equivalence of the LR-0 spectrum and $^{235}$U($n_{th}$,f) PFNS above ≈ (6 – 8) MeV is following. The SACS for reactions with the highest threshold dosimetry, namely those with $E_{50\%}$ between ≈ 5 and 15 MeV, measured in LR-0 should cause a practically negligible correction if transformed to SACS corresponding to $^{235}$U($n_{th}$,f)PFNS using a reference reactions with $E_{50\%}$ ≈ 6 - 7 MeV.

### 4.3 Spectrum averaged cross sections

Calculations of neutron and gamma transport from the core to the irradiated sample are normalized per 1 core neutron. Coupled neutron-photon transport was simulated to obtain spectrum in the reference volume, and the reaction rates of the monitors at defined positions were determined. By comparing the calculated and measured reaction rate of the monitor, a scaling factor representing the neutron emissivity of the core can be derived. The actual neutron flux in the reference volume is then obtained simply by calculating neutron flux multiplied by the scaling factor. This approach is possible as the neutron field used in the defined position of the monitors is the benchmark reference neutron field (Kostal et al., 2020), where the mathematical models used for the flux calculations were validated within study of the criticality (Kostal et al., 2016a), flux distribution (Kostal et al., 2018b), fission source distribution (Kostal et al., 2016b, Kostal et al 2022), and neutron spectrum (Kostal et al., 2015).

With the knowledge of the neutron flux in the target, the reaction rates derived by HPGe spectrometry can be easily transformed into a spectrum averaged cross section. The biggest advantage of this approach is that new experiments can supplement the primary data set by other reaction. Only in the case of capture reactions, corrections for RR (reaction rate) changes due to slight changes in the neutron spectrum (spectral shift corrections) must be applied. This correction is defined as the normalization constant between the given calculated weighted cross section and the reference value based on the reference model connected with the experiments in 2020. However, this correction is small in order of 0.1 %, but can reach to about 2 % in the case of $^{58}$Fe(n,γ) reaction and 5 % in the case of $^{58}$Fe(n,γ) in cadmium cladding. This set consists of separate reactions evaluated from small activation foils (see Chapter 4.3.1), while reactions with very high thresholds derived from large samples are evaluated in Chapter 4.3.2.

Uncertainty in reaction rates covers scaling factor uncertainty and measurement uncertainty. It considers the uncertainty of the counting rates, being lower than 1.5 %, and includes the following main components: gross peak area, Compton continuum area, background area, and propagation of uncertainties in the energy and peak shape calibrations. The detector efficiency uncertainty was derived from the difference between the experimentally determined efficiency and the efficiency determined with a precise mathematical model and is about 1.9 %.

In addition to the above uncertainties, there are also other stochastic uncertainties: the radionuclide half-life value or branching ratios. However, these uncertainties are negligible compared to the count rate uncertainties.

#### 4.3.1 Evaluation of small foils in reference volume

In this experimental set, three independent experiments have been performed with small foils with the weight of a few grams or smaller. The foils were placed in the defined spaces drilled in the pins of the foil holder and also in the center among the glass vials (see Figure 2). The summary of the experiments is given in Table 2, together with relative uncertainties (1σ level). The different critical water level $H_{cr}$ reflects different positions of the instrumentation tubes in each arrangement. The reference volume ($H_{ref}$) is also slightly shifted regarding to actual geometrical needs, with the positions of monitoring foils being the same in all experiments and the axial levels are defined relative to beginning of fission column.

The first experiment used alanine as nitrogen containing compound. It was placed in a glass vial to prevent a leak of $^{14}$C during irradiation and to minimize the distortion of the neutron field around it. All activation detectors in this arrangement were placed into the central position of the holder (see Kostal et al., 2018b). This central position was surrounded by monitoring Au, Ta, and Ni foils. Foils placed in the pins of the foil carrier were used to evaluate the flux gradient. Therefore, it can be assumed that the neutron flux in the detector and monitoring foils is close to equality.

The calculated gradient between the monitoring and central positions, determined for the $^{58}Ni(n,p)$ reaction rates, together with experimental validation, is listed in Table 3, from which it is apparent that the flux variations in reference volume are within less than 5 %. Their positions relative to the origin of the benchmark model coordinate system (the center of the dry hexagonal channel at the beginning of the fission column) are expressed by the Cartesian coordinates in Table 3.

Table 2.: Summary of experiments with small foils

| | 2020 configuration | | 2021 configuration | | 2022 configuration | |
|---|---|---|---|---|---|---|
| | Mean | Rel. unc. | Mean | Rel. unc. | Mean | Rel. unc. |
| Scaling [n/s] | 7.96E+11 | 2.3% | 6.90E+11 | 3.0% | 7.93E+11 | 2.6% |
| Flux [$cm^{-2}\cdot s^{-1}$] | 2.20E-4 | 0.1% | 2.18E-4 | 0.1% | 2.20E-4 | 0.0% |
| $H_{cr.}$ [cm] | 53.553 | 0.01% | 54.412 | 0.01% | 53.800 | 0.01% |
| $H_{ref.}$ [cm] | 21.750 | 0.05% | 21.750 | 0.05% | 22.300 | 0.04% |

Table 3.: Calculated gradient between monitoring positions and in the reference volume for the $^{58}Ni(n,p)$ reaction rates together with experimental validation during the first irradiation experiment. The coordinates are relative to the center of dry hexagonal channel, axially at beginning of the fission column

| Position | Pin 1<br>x=4.7 y=8.14 | Pin 2<br>x=4.7 y=-8.14 | Pin 3<br>x=-9.4 y=0 | Pin 4<br>x=0 y=0 |
|---|---|---|---|---|
| z=18.35 cm | 1.014 | 1.014 | 1.011 | 0.967 |
| z=23.35 cm | 1.042 | 1.047 | 1.042 | 1.042 |
| z=28.35 cm | 1.044 | 1.046 | 1.045 | 1.005 |
| z=33.35 cm | 1.006 | 1.007 | 1.004 | 0.965 |
| | | C/E-1 | | |
| z=18.35 cm | 0.6% | 0.3% | 3.0% | 0.3% |
| z=23.35 cm | - | - | - | - |
| z=28.35 cm | -2.2% | 2.0% | - | 2.4% |
| z=33.35 cm | -0.8% | -0.2% | -0.3% | -1.4% |

The spectrum averaged cross sections obtained in each experiment are listed in Table 4. The actual values are valid for the relevant spectrum. Thus, in the case of capture reactions, the spectral shift correction factor (see relative spectrum difference in Figure 3) has to be applied for direct comparison (see Table 6). However, in the case of threshold reactions, the tabulated data can be compared directly and the independently measured cross section values show very good agreement. The mean deviation is mostly below 1 %. Differences above 3 % can be found only in the case of $^{64}Zn(n,p)$ and $^{60}Ni(n,p)$.

Table 4.: Summary of evaluated integral cross sections averaged in LR-0 spectrum

| | 2020 configuration | | 2021 configuration | | 2022 configuration | |
|---|---|---|---|---|---|---|
| Reaction | Mean [b] | Rel. unc. | Mean [b] | Rel. unc. | Mean [b] | Rel. unc. |
| $^{58}$Fe(n,γ) | 2.221E-1 | 3.0% | 2.078E-1 | 4.1% | | |
| $^{59}$Co(n,γ) | | | 7.037E+0 | 4.1% | | |
| Cd $^{197}$Au(n,γ) | | | | | 3.693E+1 | 3.8% |
| Cd $^{23}$Na(n,γ) | | | | | 8.231E-3 | 3.3% |
| Cd $^{58}$Fe(n,γ) | | | | | 3.265E-2 | 3.5% |
| $^{115}$In(n,n') | | | 5.800E-2 | 4.0% | 5.848E-2 | 3.4% |
| $^{47}$Ti(n,p) | 5.008E-3 | 3.1% | 4.995E-3 | 4.0% | 4.995E-3 | 3.2% |
| $^{64}$Zn(n,p) | | | 1.014E-2 | 4.0% | 1.112E-2 | 3.4% |
| $^{58}$Ni(n,p) | 2.955E-2 | 3.0% | | | 2.949E-2 | 3.2% |
| $^{54}$Fe(n,p) | 2.191E-2 | 3.5% | | | 2.167E-2 | 3.2% |
| $^{92}$Mo(n,p)$^{92m}$Nb | 1.928E-3 | 3.1% | 1.949E-3 | 4.1% | 1.914E-3 | 3.4% |
| $^{46}$Ti(n,p) | 2.981E-3 | 3.0% | | | 2.984E-3 | 3.2% |
| $^{60}$Ni(n,p) | | | 5.393E-4 | 7.2% | 6.214E-4 | 5.9% |
| $^{63}$Cu(n,α) | 1.405E-4 | 3.5% | | | 1.382E-4 | 4.2% |
| $^{54}$Fe(n,α) | | | | | 2.423E-4 | 10.5% |
| $^{56}$Fe(n,p) | 2.974E-4 | 3.1% | 2.864E-4 | 4.5% | 2.926E-4 | 3.5% |
| $^{48}$Ti(n,p) | 8.070E-5 | 3.2% | 7.982E-5 | 5.0% | 8.230E-5 | 3.2% |
| $^{24}$Mg(n,p) | | | | | 3.929E-4 | 4.6% |
| $^{27}$Al(n,α) | 1.873E-4 | 3.6% | 1.872E-4 | 4.3% | 1.902E-4 | 3.4% |
| $^{51}$V(n,α) | 6.345E-6 | 3.3% | | | 6.652E-6 | 3.9% |
| $^{197}$Au(n,2n) | 9.382E-4 | 4.0% | | | | |
| $^{93}$Nb(n,2n)$^{92m}$Nb | 1.165E-4 | 3.3% | 1.197E-4 | 4.1% | 1.233E-4 | 3.5% |
| $^{55}$Mn(n,2n) | | | | | 6.466E-5 | 4.5% |
| $^{58}$Ni(n,x)$^{57}$Co | 6.30E-05 | 13% | 6.30E-05 | 6% | 6.81E-05 | 6% |

### 4.3.2 Evaluation of large volume samples

The previous chapter refers to the use of small foils. In a low flux reference field, the activation of foils with a higher threshold reaction may be unmeasurable. This is why larger foils or samples are sometimes used.

In that case, the monitoring Al and Ni foils are placed directly on the casing with the irradiated material. Monitoring foils allow determination of the core emissivity during the experiment and the related flux in the sample. A direct comparison with the previous case is not possible.

Due to very high thresholds, mostly above 10 MeV, where the transport cross sections have a flat character, the spectral shift is not a problem, so the spectrum in the target is identical to the LR-0 spectrum above this threshold. The only significant effect affecting the reaction rates is the flux loss (see Kostal et al 2017b). Considering this effect, data for large targets can be linked with data for small targets. The reaction rates data in large volume samples are listed in Table 5.

In the first experiment with large set of foil, which is used as a reference, the neutron flux above 10 MeV per 1 core neutron was determined to be 8.394E-8 $cm^{-2}\cdot s^{-1}$. As expected, the flux in the various targets is lower, see Table 5. The above-mentioned flux loss is proportional to the ratio of the flux in the void case to the flux in the material target. The largest flux loss can be observed in the case of $As_2O_3$. This reflects the high amount and high density of used arsenic oxide.

Table 5.: Summary of older experimental results normalized to set to experimental data from 2020 (Table 4)

| | Critical water level [cm] | RR [$s^{-1}$] | Scaling [n/s] | Normalized flux in target $E_n$ > 10 MeV [$cm^{-2}$] | Rel. unc. |
|---|---|---|---|---|---|
| $^{127}$I(n,2n) | 56.08 | 2.785E-20 | 4.199E+11 | 7.727E-8 | 4.0% |
| $^{75}$As(n,2n) | 57.36 | 6.395E-21 | 4.014E+11 | 6.769E-8 | 4.3% |
| $^{89}$Y(n,2n) | 56.50 | 3.777E-21 | 4.037E+11 | 7.558E-8 | 3.2% |
| $^{19}$F(n,2n) | 55.65 | 3.034E-22 | 7.316E+11 | 7.399E-8 | 4.0% |
| $^{90}$Zr(n,2n) | 56.19 | 2.403E-21 | 3.821E+11 | 8.189E-8 | 4.0% |
| $^{23}$Na(n,2n) | 57.20 | 9.327E-23 | 4.516E+11 | 7.182E-8 | 4.8% |

### 4.3.3 Evaluation of reaction rates, SACS and comparison with calculations

Based on the result for the small and large samples, an extensive set of reaction rates was compiled, which can be found in Table 6. One of the main advantages of this data set its applicability also for testing of the energy spectrum deconvolution codes, since the set includes many reactions covering a wide energy range from thermal to 16 MeV. C/E comparisons of MCNP 6.2 and TRIPOLI calculation are presented in Table 7. It is worth noting that the agreement reached with both codes and IRDFF-II library is good for lower energy threshold reactions. For higher threshold reactions the discrepancies resulting from the incorrect representation of the $^{235}$U($n_{th}$,f) PFNS in JEFF-3.1.1 can be seen.

A comparison with MCNP 6.2 with ENDF/B-VIII.0 and the TRIPOLI with JEFF-3.1.1 calculation is shown in Table 7. It is worth noting that the agreement obtained with both codes and the IRDFF-II library is good for reactions with a lower energy threshold. Inconsistencies resulting from the incorrect representation of the $^{235}$U($n_{th}$,f) PFNS in JEFF-3.1.1 can be seen for higher threshold responses.

Table 6.: SACS averaged in LR-0 spectrum evaluated using the calculated neutron flux approach.

| Reaction | Mean [b] | Unc. |
|---|---|---|
| $^{58}$Fe(n,γ) | 2.150E-1 | 4.2% |
| $^{59}$Co(n,γ) | 7.037E+0 | 4.1% |
| Cd $^{197}$Au(n,γ) | 3.693E+1 | 3.8% |
| Cd $^{23}$Na(n,γ) | 8.231E-3 | 3.3% |
| Cd $^{58}$Fe(n,γ) | 3.265E-2 | 3.5% |
| $^{115}$In(n,n') | 5.824E-2 | 2.7% |
| $^{47}$Ti(n,p) | 4.999E-3 | 2.0% |
| $^{64}$Zn(n,p) | 1.063E-2 | 5.3% |
| $^{58}$Ni(n,p) | 2.952E-2 | 2.2% |
| $^{54}$Fe(n,p) | 2.179E-2 | 2.4% |
| $^{92}$Mo(n,p)$^{92m}$Nb | 1.930E-3 | 2.2% |
| $^{46}$Ti(n,p) | 2.983E-3 | 2.2% |
| $^{60}$Ni(n,p) | 5.803E-4 | 8.5% |
| $^{63}$Cu(n,α) | 1.394E-4 | 2.9% |
| $^{54}$Fe(n,α) | 2.423E-4 | 10.5% |
| $^{56}$Fe(n,p) | 2.921E-4 | 2.6% |
| $^{48}$Ti(n,p) | 8.094E-5 | 2.6% |
| $^{24}$Mg(n,p) | 3.929E-4 | 4.6% |
| $^{27}$Al(n,α) | 1.882E-4 | 2.3% |
| $^{51}$V(n,α) | 6.498E-6 | 3.5% |
| $^{197}$Au(n,2n) | 9.382E-4 | 4.0% |
| $^{93}$Nb(n,2n)$^{92m}$Nb | 1.198E-4 | 3.1% |
| $^{127}$I(n,2n) | 3.276E-4 | 4.0% |
| $^{55}$Mn(n,2n) | 6.466E-5 | 4.5% |
| $^{75}$As(n,2n) | 8.982E-5 | 4.3% |
| $^{89}$Y(n,2n) | 4.724E-5 | 3.2% |
| $^{19}$F(n,2n) | 2.139E-6 | 4.0% |
| $^{90}$Zr(n,2n) | 2.931E-5 | 4.0% |
| $^{23}$Na(n,2n) | 1.097E-6 | 4.8% |
| $^{58}$Ni(n,x)$^{57}$Co | 6.469E-5 | 6.4% |

In reactions marked Cd foils were covered by Cd layer.

Table 7.: C/E-1 of SACS averaged in LR-0 spectrum compared with various calculations.

| | MCNP 6.2 | | TRIPOLI | |
|---|---|---|---|---|
| | IRDFF-II | ENDF/B-VIII.0 | IRDFF-II | JEFF-3.1.1 |
| $^{58}$Fe(n,γ) | 2.9% | 3.2% | 1.0% | 0.9% |
| $^{59}$Co(n,γ) | 4.6% | 4.6% | 3.1% | 3.4% |
| Cd $^{197}$Au(n,γ) | -1.1% | - | -0.7% | 0.2% |
| Cd $^{23}$Na(n,γ) | 5.0% | 5.4% | 5.1% | 3.3% |
| Cd $^{58}$Fe(n,γ) | 4.4% | 6.1% | 7.7% | 7.7% |
| $^{115}$In(n,n') | -4.0% | - | -3.2% | - |
| $^{47}$Ti(n,p) | 1.6% | 7.3% | 2.6% | 9.3% |
| $^{64}$Zn(n,p) | -0.1% | -9.5% | 0.9% | 3.5% |
| $^{58}$Ni(n,p) | 0.1% | -1.0% | 1.2% | 1.6% |
| $^{54}$Fe(n,p) | -1.9% | -1.9% | -0.7% | -10.1% |
| $^{92}$Mo(n,p)$^{92m}$Nb | -6.1% | - | -4.3% | - |
| $^{46}$Ti(n,p) | 2.3% | -6.2% | 4.7% | 1.7% |
| $^{60}$Ni(n,p) | -0.1% | -7.1% | 2.8% | 3.0% |
| $^{63}$Cu(n,α) | 1.6% | 45.4% | 4.4% | 2.1% |
| $^{54}$Fe(n,α) | 5.7% | -16.2% | 8.1% | -42.4% |
| $^{56}$Fe(n,p) | -0.3% | -0.3% | 2.7% | 0.9% |
| $^{48}$Ti(n,p) | -0.2% | 17.1% | 2.2% | -8.2% |
| $^{24}$Mg(n,p) | 1.6% | 10.6% | 4.3% | 13.7% |
| $^{27}$Al(n,α) | 0.4% | 2.3% | 2.9% | 4.8% |
| $^{51}$V(n,α) | 1.4% | 8.7% | 2.1% | 2.3% |
| $^{197}$Au(n,2n) | -2.9% | -7.6% | -2.8% | -7.7% |
| $^{93}$Nb(n,2n)$^{92m}$Nb | 4.2% | - | 3.0% | - |
| $^{127}$I(n,2n) | 0.1% | 1.5% | -1.7% | 8.7% |
| $^{55}$Mn(n,2n) | 6.7% | 4.5% | 0.4% | 2.7% |
| $^{75}$As(n,2n) | 0.6% | 3.5% | -5.4% | -12.0% |
| $^{89}$Y(n, 2n) | 2.1% | 3.2% | -7.5% | -8.0% |
| $^{19}$F(n,2n) | 5.6% | 26.6% | -4.9% | 14.4% |
| $^{90}$Zr(n,2n) | 0.9% | 2.9% | -10.7% | -5.0% |
| $^{23}$Na(n,2n) | -1.5% | 55.2% | -16.6% | -17.4% |
| $^{58}$Ni(n,x)$^{57}$Co | - | -4.4% | - | -9.1% |

In reactions marked Cd foils were covered by Cd layer.

The spectrum average neutron induced cross sections were also calculated by code RR_UNC (Trkov et al., 2001). The neutron spectrum was simulated by MCNP6.2 with ENDF/B-VIII.0 data, see section 4.2 and Figure 4. The neutron reaction cross sections and their covariances were taken from the IRDFF-II library (Trkov et al., 2020). Exception was the reaction $^{58}$Ni(n,x)$^{57}$Co, not included in IRDFF-II, for which the cross sections were taken from ENDF/B-VIII.0. Besides the main production channel $^{58}$Ni(n,np+d)$^{57}$Co, the pathway $^{58}$Ni(n,2n)$^{57}$Ni followed by the EC or $\beta^+$ decay to $^{57}$Co was included in calculations since the $^{57}$Ni half-life $T_{1/2}$ = 35.6 h is larger than irradiation and cooling times used in measurement. The LR-0 neutron spectrum and all reaction cross sections were processed into the 725 energy groups presentation by PREPRO-2023 (Cullen 2023) to feed RR_UNC.

The experimental and calculated neutron induced SACS for the 27 reactions benchmarked in the in-core dry hexagonal channel of LR-0 are summarized in Table 8, and as C/E ratio are graphically compared in Figure 7. The absolute values and uncertainties for experimental SACS are copied from Table 6. For all reactions studied, we observe an agreement within the estimated experimental and calculated uncertainties. It is worth noting that the SACS uncertainties resulting from the MCNP simulation of the LR-0 spectra are smaller than those propagated from the IRDFF-II cross sections, even for the high threshold reactions.

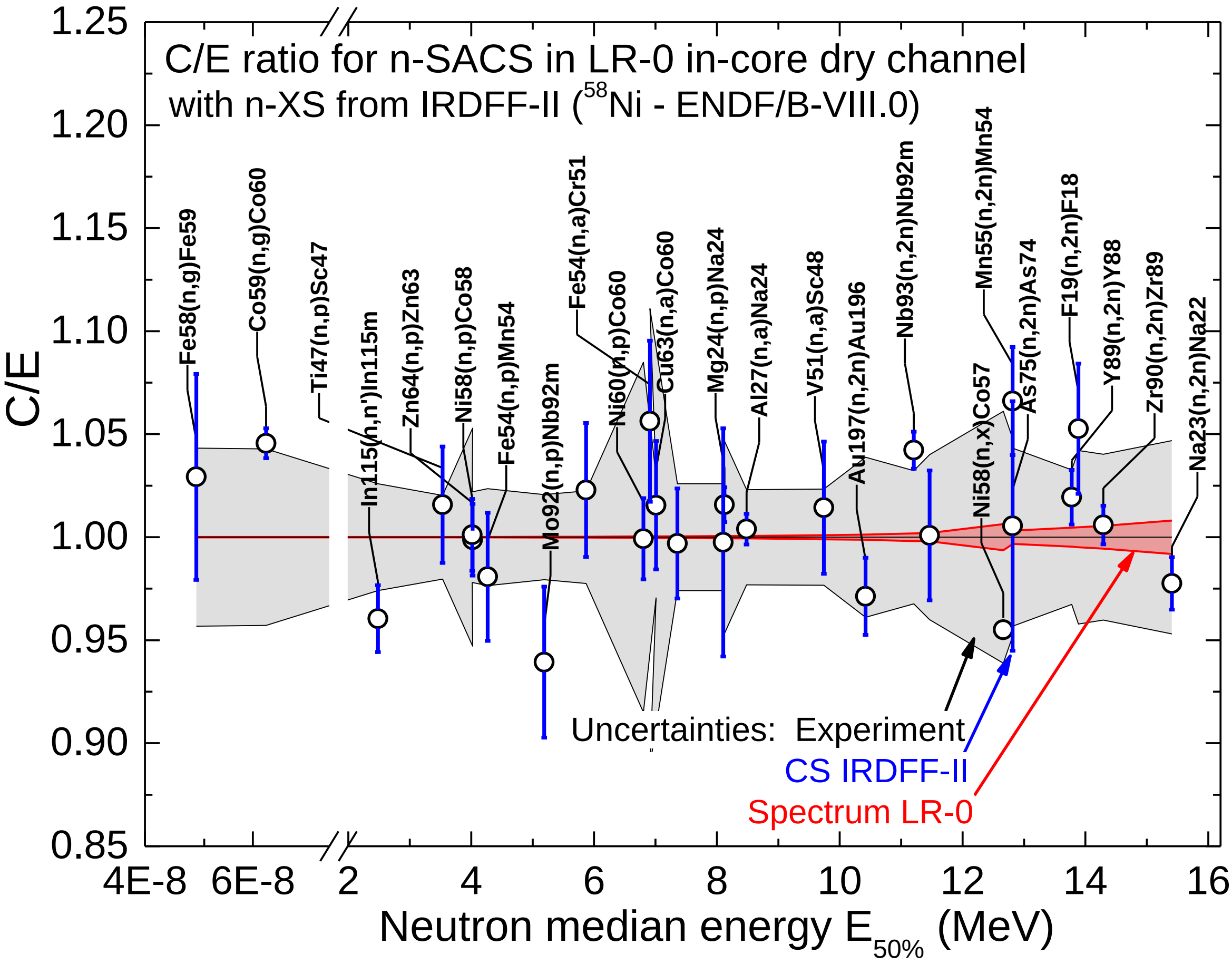

Figure 7.: C/E ratio for the neutron induced spectrum average cross sections (n-SACS) versus the neutron median energy $E_{50\%}$. The gray corridor area shows the total experimental uncertainties, the blue columns – the total uncertainties from IRDFF-II, the red area – the statistics of the simulated neutron spectrum. Note the change in scale of the abscissas at neutron energies below 2 MeV.

Table 8.: SACS and their uncertainties measured in the dry hexagonal channel of the LR-0 reactor.

| Reaction | $E_{50\%}$ [MeV] | Experiment | | Calculation | | | | Ratio | |
|---|---|---|---|---|---|---|---|---|---|
| | | SACS [mb] | Unc. [%] | n-SACS | Un(XS) [%] | Un(Sp) [%] | Un(Tot) [%] | C/E | Δ(C/E) [%] |
| $^{58}Fe(n,\gamma)^{59}Fe$ | 4.854E-08 | 2.150E+02 | 4.20 | 2.213E+02 | 4.86 | 0.01 | 4.86 | 1.029 | 6.42 |
| $^{59}Co(n,\gamma)^{60}Co$ | 6.307E-08 | 7.037E+03 | 4.10 | 7.358E+03 | 0.69 | 0.01 | 0.69 | 1.046 | 4.16 |
| $^{115}In(n,n')^{115m}In$ | 2.481 | 5.824E+01 | 2.70 | 5.594E+01 | 1.69 | 0.01 | 1.69 | 0.960 | 3.19 |
| $^{47}Ti(n,p)^{47}Sc$ | 3.534 | 4.999E+00 | 2.00 | 5.078E+00 | 2.78 | 0.01 | 2.78 | 1.016 | 3.42 |
| $^{64}Zn(n,p)^{64}Cu$ | 4.022 | 1.063E+01 | 5.30 | 1.062E+01 | 1.74 | 0.01 | 1.74 | 0.999 | 5.58 |
| $^{58}Ni(n,p)^{58}Co$ | 4.018 | 2.952E+01 | 2.20 | 2.955E+01 | 1.74 | 0.01 | 1.74 | 1.001 | 2.80 |
| $^{54}Fe(n,p)^{54}Mn$ | 4.267 | 2.179E+01 | 2.40 | 2.137E+01 | 3.17 | 0.01 | 3.17 | 0.981 | 3.98 |
| $^{92}Mo(n,p)^{92m}Nb$ | 5.187 | 1.930E+00 | 2.20 | 1.813E+00 | 3.90 | 0.02 | 3.90 | 0.939 | 4.48 |
| $^{46}Ti(n,p)^{46}Sc$ | 5.869 | 2.983E+00 | 2.20 | 3.051E+00 | 3.18 | 0.02 | 3.18 | 1.023 | 3.87 |
| $^{60}Ni(n,p)^{60}Co$ | 6.803 | 5.803E-01 | 8.50 | 5.799E-01 | 1.97 | 0.04 | 1.98 | 0.999 | 8.73 |
| $^{63}Cu(n,\alpha)^{60}Co$ | 7.008 | 1.394E-01 | 2.90 | 1.416E-01 | 3.07 | 0.04 | 3.07 | 1.015 | 4.22 |
| $^{54}Fe(n,\alpha)^{51}Cr$ | 6.909 | 2.423E-01 | 10.50 | 2.560E-01 | 3.70 | 0.04 | 3.70 | 1.056 | 11.13 |
| $^{56}Fe(n,p)^{56}Mn$ | 7.355 | 2.921E-01 | 2.60 | 2.912E-01 | 2.68 | 0.04 | 2.68 | 0.997 | 3.73 |
| $^{48}Ti(n,p)^{48}Sc$ | 8.103 | 8.094E-02 | 2.60 | 8.073E-02 | 5.55 | 0.06 | 5.55 | 0.997 | 6.13 |
| $^{24}Mg(n,p)^{24}Na$ | 8.122 | 3.929E-01 | 4.60 | 3.991E-01 | 0.83 | 0.06 | 0.83 | 1.016 | 4.67 |
| $^{27}Al(n,\alpha)^{24}Na$ | 8.482 | 1.882E-01 | 2.30 | 1.889E-01 | 0.74 | 0.06 | 0.74 | 1.004 | 2.42 |
| $^{51}V(n,\alpha)^{48}Sc$ | 9.741 | 6.498E-03 | 2.30 | 6.591E-03 | 3.16 | 0.10 | 3.16 | 1.014 | 3.91 |
| $^{197}Au(n,2n)^{196}Au$ | 10.422 | 9.382E-01 | 4.00 | 9.113E-01 | 1.93 | 0.13 | 1.93 | 0.971 | 4.44 |
| $^{93}Nb(n,2n)^{92m}Nb$ | 11.206 | 1.198E-01 | 3.10 | 1.249E-01 | 0.86 | 0.18 | 0.88 | 1.042 | 3.22 |
| $^{127}I(n,2n)^{126}I$ | 11.462 | 3.276E-01 | 4.00 | 3.279E-01 | 3.15 | 0.20 | 3.16 | 1.001 | 5.10 |
| $^{58}Ni(n,x)^{57}Co$ | 12.663 | 6.469E-02 | 6.40 | 6.178E-02 | 0.00 | 0.66 | 0.27 | 0.955 | 6.41 |
| $^{55}Mn(n,2n)^{54}Mn$ | 12.814 | 6.466E-02 | 4.50 | 6.893E-02 | 2.46 | 0.31 | 2.48 | 1.066 | 5.14 |
| $^{75}As(n,2n)^{74}As$ | 12.816 | 8.982E-02 | 4.30 | 9.031E-02 | 6.02 | 0.32 | 6.03 | 1.005 | 7.41 |
| $^{89}Y(n,2n)^{88}Y$ | 13.778 | 4.724E-02 | 3.20 | 4.816E-02 | 1.30 | 0.45 | 1.37 | 1.019 | 3.48 |
| $^{19}F(n,2n)^{18}F$ | 13.885 | 2.139E-03 | 4.00 | 2.252E-03 | 3.00 | 0.46 | 3.04 | 1.053 | 5.02 |
| $^{90}Zr(n,2n)^{88}Zr$ | 14.291 | 2.931E-02 | 4.00 | 2.948E-02 | 0.93 | 0.55 | 1.08 | 1.006 | 4.14 |
| $^{23}Na(n,2n)^{24}Na$ | 15.410 | 1.097E-03 | 4.80 | 1.072E-03 | 1.30 | 0.83 | 1.54 | 0.978 | 5.04 |

The neutron induced n-SACS, uncertainty components Un (Un(XS) is 1σ uncertainty of cross section, Un(Sp) is 1σ uncertainty of neutron spectrum, and Un(Tot) is total uncertainty) and median energy $E_{50\%}$ were calculated with cross sections from IRDFF-II, except $^{28}Ni(nx,x)^{57}Co$ - from ENDF/B-VIII.0. The calculation over experiment ratio C/E and total uncertainty Δ(C/E) are given in last two columns.

## 4.4 Correction to sections $^{235}U(n_{th},f)$ PFNS

Thanks to the similarity of the LR-0 spectrum with $^{235}U(n_{th},f)$ PFNS in the region above 6 MeV, the SACS averaged in LR-0 spectrum with a threshold above 6 MeV can be evaluated as SACS averaged in $^{235}U(n_{th},f)$ PFNS (see Eq. 3). It means that if the threshold reaction with a threshold over 6 MeV is measured, the interacting part of the neutron spectrum has the same shape as $^{235}U(n_{th}, f)$ PFNS.

In the evaluated ENDF/B-VIII.0 $^{235}U(n_{th},f)$ PFNS, about 2.566% of the total emitted neutrons have energy above 6 MeV. In the LR-0 spectrum, only 0.714 % of neutrons have energy above 6 MeV. Thus, for the SACS determination, the flux used for normalization of the calculated RR is smaller by a factor of 3.5942, reflecting that a large amount of thermal and epithermal neutrons is added to the "true PFNS" part. In other words, this factor is the ratio between the area under the curve of PFNS and LR-0 one as in Figure 8, and it is also the multiplication factor to convert SACS averaged in LR-0 into SACS averaged in $^{235}U(n_{th},f)$ PFNS. The resulting spectrum averaged cross sections (SACS) in $^{235}U(n_{th},f)$ PFNS, obtained using (3), are listed in Figure 9.

The correction was applied to all threshold reactions to also test the reactions with a threshold below 6 MeV. It is worth noting that despite the lower IRDFF-II energy threshold, the cross sections are in very good agreement with the measured values. This reflects the fact that there are oscillations between the actual reactor and $^{235}U(n_{th},f)$ PFNS spectra, but on average both spectra are the same above 6 MeV. A notable discrepancy can be found in the case of $^{115}In(n,n')$, where the calculation is significantly smaller than the IRDFF-II value. This is consistent with the assumption that the thermal reactor spectrum is much higher than $^{235}U(n_{th},f)$ PFNS in the sensitive region for this reaction.

The last row shows the result of the cross sections of the $^{58}Ni(n,x)^{57}Co$ reaction. This reaction is not in IRDFF-II, so it was compared with the ENDF/B-VIII.0 data. As this reaction is potentially a dosimetry reaction, it was also compared with previously measured values (see Table 10). It is worth noting that the agreement is quite good, the average value excluding the outlier differs from the current result by 1 %, so this reaction can be recommended as dosimetry reaction.

$$\overline{\sigma}^{exp} = \frac{1}{K} \times \frac{q(\overline{P})}{\int \phi(E) \cdot dE} \times R \qquad (3)$$

Where:

$\overline{\sigma}^{exp}$; is the $^{235}U(n_{_th},$ fiss.) PFNS averaged cross section;

$K$; is the scaling factor based on absolute flux density, neutron emission per second;

$q(\overline{P})$; is the experimentally measured reaction rate (see Equation (2));

$\phi(E)$; is the calculated neutron spectrum normalized per 1 neutron in core;

$R$; is the ratio between share of neutrons with energy above 6 MeV in $^{235}U(n_{th},f)$ PFNS (2.566% in ENDF/B-VIII.0) and LR-0 spectra (0.714%) being 3.5942.

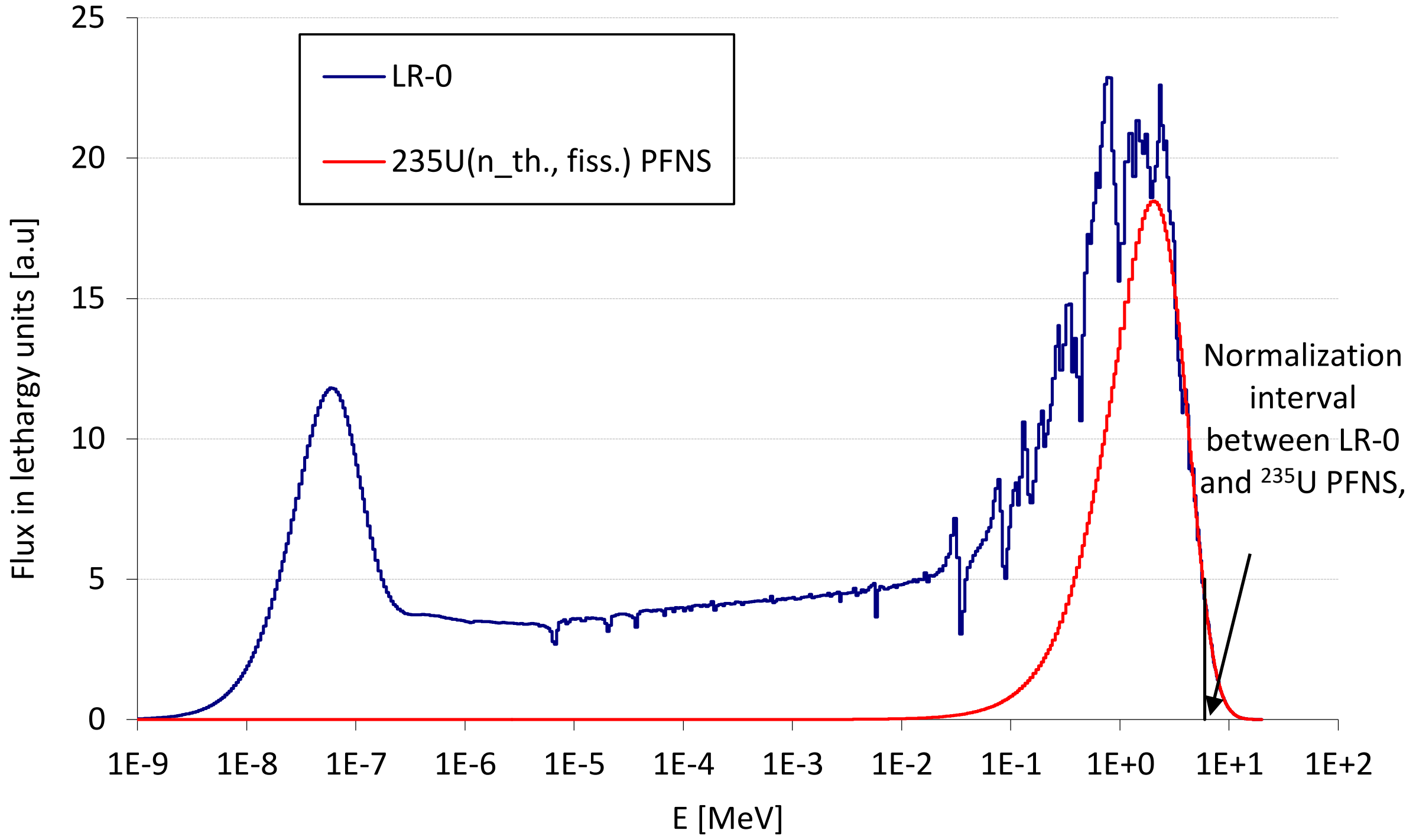

Figure 8.: Graphical interpretation of similarity in LR-0 spectrum and $^{235}U(n_{th},f)$ PFNS and energy interval where they were normalized to each other.

Table 9.: Summary of SACS corrected to $^{235}U(n_{th}, f)$ PFNS and compared with IRDFF-II data taken from [Trkov et al., 2020]

| | Threshold energy [MeV] | SACS [mb] | Rel. unc. | Eval./E-1 |
|---|---|---|---|---|
| $^{115}In(n,n')$ | 0 | 209.3 | 2.7% | -10.3% |
| $^{47}Ti(n,p)$ | 0 | 17.97 | 2.0% | -0.7% |
| $^{64}Zn(n,p)$ | 0 | 38.21 | 5.3% | 1.8% |
| $^{58}Ni(n,p)$ | 0 | 106.1 | 2.2% | 2.0% |
| $^{54}Fe(n,p)$ | 0 | 78.33 | 2.4% | -0.3% |
| $^{92}Mo(n,p)$ | 0 | 6.938 | 2.2% | -3.6% |
| $^{46}Ti(n,p)$ | 1.619 | 10.72 | 2.2% | 7.4% |
| $^{60}Ni(n,p)$ | 2.075 | 2.086 | 8.5% | 4.5% |
| $^{63}Cu(n,\alpha)$ | 0 | 0.5009 | 2.9% | 3.3% |
| $^{54}Fe(n,\alpha)$ | 0 | 0.8707 | 10.5% | -0.7% |
| $^{56}Fe(n,p)$ | 2.966 | 1.050 | 2.6% | 2.8% |
| $^{48}Ti(n,p)$ | 3.274 | 0.2909 | 2.6% | 3.6% |
| $^{24}Mg(n,p)$ | 4.932 | 1.412 | 4.6% | 2.6% |
| $^{27}Al(n,\alpha)$ | 3.25 | 0.6764 | 2.3% | 3.6% |
| $^{51}V(n,\alpha)$ | 2.093 | 0.0234 | 3.5% | 4.0% |
| $^{197}Au(n,2n)$ | 8.114 | 3.372 | 4.0% | 0.4% |
| $^{93}Nb(n,2n)$ | 9.064 | 0.4307 | 3.1% | 0.9% |
| $^{127}I(n,2n)$ | 9.217 | 1.177 | 4.0% | 1.8% |
| $^{55}Mn(n,2n)$ | 10.414 | 0.2324 | 4.5% | 0.0% |
| $^{75}As(n,2n)$ | 10.384 | 0.3228 | 4.3% | -1.1% |
| $^{89}Y(n,2n)$ | 11.611 | 0.1698 | 3.2% | 0.8% |
| $^{19}F(n,2n)$ | 10.986 | 0.00769 | 4.0% | 5.9% |
| $^{90}Zr(n,2n)$ | 12.1 | 0.1053 | 4.0% | -0.7% |
| $^{23}Na(n,2n)$ | 12.965 | 0.00394 | 4.8% | -1.9% |
| $^{58}Ni(n,x)^{57}Co$ | 6.051 | 0.2325 | 6.4% | 2.5% [a] |

[a] this reaction is not in IRDFF-II, thus data are from ENDF/B-VIII.0 (Brown et al., 2018)

Table 10.: Comparison between actually measured $^{58}Ni(n,x)^{57}Co$ SACS averaged in $^{235}U(n_{th.}; f)$ and previously measured results

| Reference | Mean [mb] | Eval./E-1 |
|---|---|---|
| BRUGGEMAN et al., 1974 | 0.216 ± 0.005 | -7.1 % |
| WÖLFLE et al., 1980 | 0.240 ± 0.035 | 3.2 % |
| HORIBE et al., 1992 | 0.232 ± 0.005 | -0.2 % |
| ZAIDI et al., 1993 | 0.253 ± 0.015 | 8.8 % |
| Arribére et al., 2001 | 0.275 ± 0.015 | 18.3 % |
| Burianova et al., 2019 | 0.239 ± 0.013 | 2.8 % |
| Kostal et al 2021 | 0.241 ± 0.015 | 3.7 % |
| Kostal et al. 2022b | 0.226 ± 0.010 | -2.9 % |

## 4.5 Competition between the gamma and neutron induced reactions in the dry hexagonal channel

The γ-ray energy spectrum at the dosimetry foil irradiation position was simulated by the MCNP6.2 code employing the ENDF/B-VIII.0 neutron-photon transport data, i.e. the same data as for the neutron spectrum calculations, see section 4.2. The γ-ray spectrum calculated in the energy interval 1.0 keV to 21.5 MeV at the sample location point inside the dry hexagonal channel of LR-0 is shown in Figure 9. The γ-ray energies $\gamma\text{-}E_{50}\%$, where the γ-SACS integrals reach 50%, are indicated by vertical bars for ten γ-induced reactions under study. Note a difference between $^{90}Zr(\gamma,n)^{89}Zr$ (MT =

201) and $^{90}Zr(\gamma,n)^{89}Zr$ (ZA = 4089) near and below the kinematic threshold, where they should be identical. The statistical uncertainties resulting from the Monte Carlo simulation are quite small. Even around 20 MeV, where they do not exceed ≈ 10 % (they are displayed as a corridor in Figure 9). The spectrum decreases exponentially with increasing γ-ray energy. The prominent discrete γ-lines observed between 7 and 10 MeV are primary γ-ray transitions to the ground states from thermal neutron capture on the iron isotopes $^{56}Fe$, $^{54}Fe$, and $^{57}Fe$. At higher energies, from 10 up to 20 MeV, the relatively smooth energy spectrum of γ-rays is mostly stem from caused by neutron induced fission of U isotopes and high energy neutron inelastic scattering (the detailed analysis of these issues but for the case of the VR-1 reactor is given in (Kostal et al. 2022b).

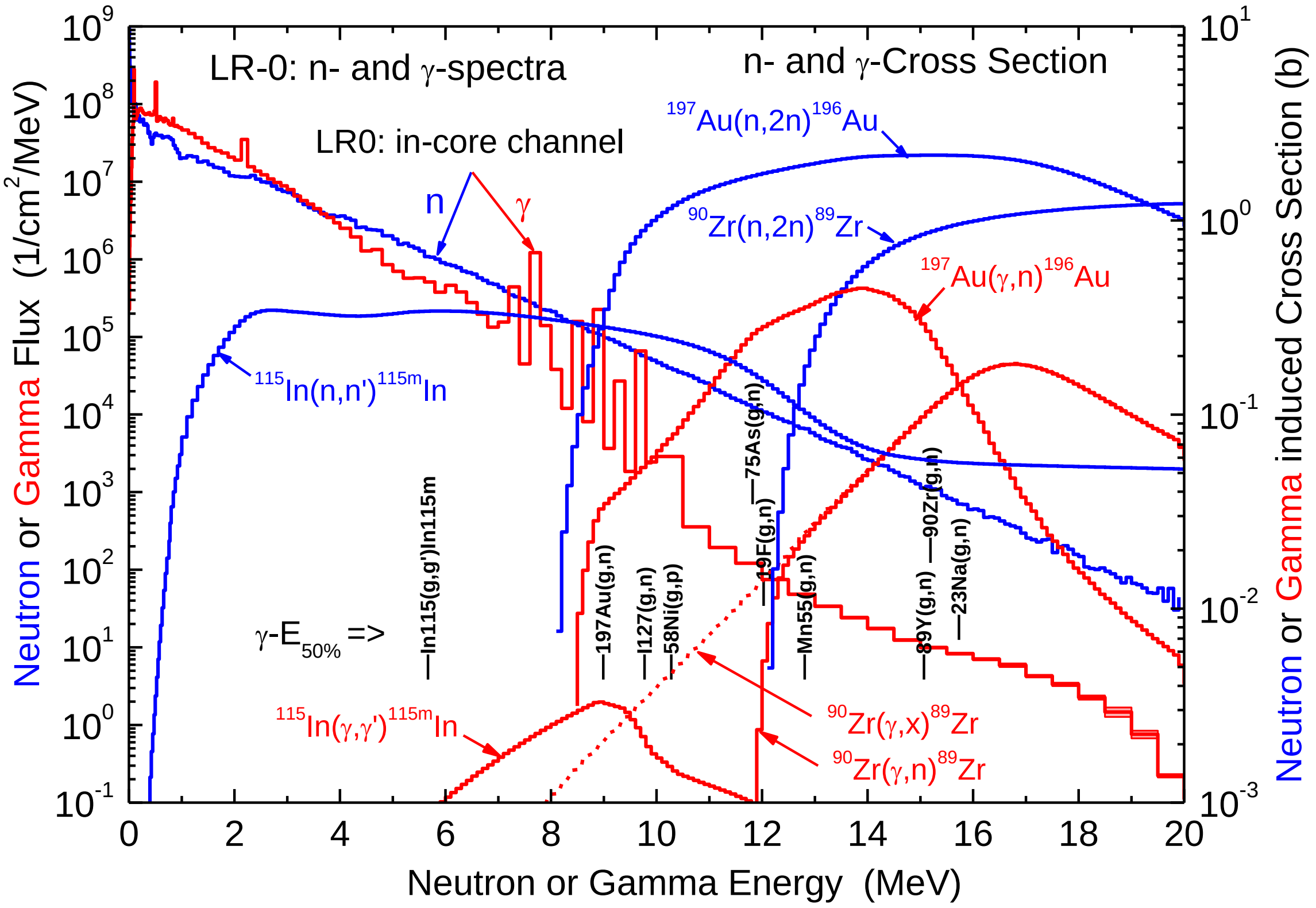

Figure 9.: MCNP calculated spectra of γ-rays. Left axis: MCNP calculated spectra of γ-rays (red curve) and neutrons (blue curve) in the dry hexagonal channel of LR-0. Right axis: cross sections for several neutron dosimetry reactions from IRDFF-II (blue curves) and for photo-nuclear reactions from IAEA/PD-2019.2 (red curves).

Gamma induced spectrum average cross sections (γ-SACS) were calculated similarly as to n-SACS using the RR_UNC code (Trkov et al., 2001). The (γ,x) cross sections were taken from the latest version of the dedicated photo-nuclear evaluated data library IAEA/PD-2019.2 (Kawano et al. 2020). The γ-ray spectrum of LR-0 and all (γ,x) reaction cross sections were processed into 725 energy groups by PREPRO-2023 (Cullen 2023). Normalizations of the γ-ray and neutron fluxes as well as the corresponding reaction rates were performed per one fission neutron in the LR-0 fissile core (i.e., per one neutron in the MCNP kcode regime).

Competing production of the resultant isotope was considered between the following pathways: (γ,γ′) and (n,n′) for one reaction pair, (γ,n) and (n,2n) for nine reaction pairs, (γ,p) and (n,np+d) for one reaction pair, $(\gamma,^{3}He)$ and (n,α) for four reaction pairs, as listed in Table 11. To illustrate the relation between the gamma and neutron cross sections, Figure 9 shows these, as an example for competing pairs (n,2n) and (γ,n) pair for the $^{90}Zr$ and $^{167}Au$ target nuclei, and (n,n′) and (γ,γ′) pair for $^{115}In$. It can be seen that the maximum values of (γ,n) cross sections are by a factor ≈ 3 – 5 lower than (n,2n), while (γ,γ′) are by ≈ 2 orders of magnitude lower than (n,n′).

The calculated γ-SACS and n-SACS and the corresponding median energies $E_{50\%}$ are listed in Table 11 together with the reaction kinematic thresholds $E_{thr}$. The impact or contribution fraction of (γ,x) is calculated as the ratio of production yields = (γ-Reaction Rate) / (n-Reaction Rate) = (γ-Flux × γ-SACS) / (n-Flux × n-SACS), where γ-Flux = 1.533 $10^{-4}$ γ/cm$^2$/s and n-Flux = 3.017 $10^{-4}$ γ/cm$^2$/s was found in the MCNP criticality simulation per one kcode neutron.

As seen in Table 11 and Figure 10, the high threshold neutron dosimetry reactions (n,2n) show the maximum competition from the photo-nuclear reactions (γ,n). The maximum contribution of around 1 % is observed for $^{197}$Au target and is less than 0.3 % for the other targets. Gamma induced impact for other reaction pairs, (γ,p) vs. (n,np+d), (γ,γ') vs. (n,n') and (γ,$^3$He) vs. (n,α) is equal or lesser than 0.1 %, see Table 11. The competition of (γ,$^3$He)/(n,α) on the four targets $^{27}$Al, $^{54}$Fe, $^{51}$V and $^{63}$Cu is extraordinarily small and presently cannot be reliably estimated due to the extreme high gamma-induced reaction threshold $E_{thr}$(γ,$^3$He) ≈ 18.9 - 23.7 MeV.

Figure 10 also shows the absolute ratio of gamma and neutron spectra for energies above 5 MeV at irradiation point. This ratio is normally less than unity except in a few energy bins where the number of gammas exceeds the number of neutrons. Visually inspecting Figure 10, we conclude that the maximum (γ,x) contributions occurs when the gamma induced reaction threshold is slightly lower than the discrete energy from ($n_{th}$,$\gamma_0$) on one of the Fe stable isotopes. Such correlation is especially clear for (γ,n) and (n,2n) pairs on $^{197}$Au, $^{93}$Nb, and $^{127}$I targets, with Au being the most sensitive to gammas with the (γ,n)/(n,2n) fraction being close to 1 %. The median energy γ-$E_{50\%}$ for these reactions is located closer to the left boundary of the (5 – 95) % reaction rate response domain (shown as horizontal bars in Figure 10).

It is interesting to note that a similar dominance effect of photo-over neutron-induced reactions on gold was observed in the VR-1 reactor. In this environment, the effect reached ≈ 50 % outside the fission core, where the neutron flux decreased stronger than γ-flux, see (Kostal et al. 2022b). Inside the VR-1 core, fraction of (γ,p)/(n,2n) on $^{197}$Au was estimated to be 0.61 %, a value comparable to that obtained inside LR-0.

For the target isotope $^{90}$Zr, we found that the IAEA/PD-2019.2 library has a significant difference between the cross sections for $^{90}$Zr(γ,n)$^{89}$Zr (ENDF reaction number MT = 201) and $^{90}$Zr(γ,x)$^{89}$Zr (final product identification ZA = 40089g) at γ-ray energy in near the kinematic threshold, where they should be identical. The activation reaction $^{90}$Zr(γ,x)$^{89}$Zr, which should be selected for the γ-SACS calculation as correctly identifying the final product "ZAg = 40089g", wrongly extends below the kinematic threshold $E_{thr}$ = 11.98 MeV, see Figure 9. Selection of the $^{90}$Zr(γ,x)$^{89}$Zr cross section results in to extremely high, around 10 %, contribution of (γ,n) on $^{90}$Zr. To avoid this obvious defect of IAEA/PD-2019.2, we selected the cross section for reaction $^{90}$Zr(γ,n)$^{89}$Zr (MT = 201) for our γ-SACS calculation. The latter is however the neutron production reaction (γ,1n) + 2×(γ,2n) + ... , but it is identical to $^{90}$Zr(γ,x)$^{89}$Zr at energies up to the threshold 21.3 MeV for $^{90}$Zr(γ,2n)$^{88}$Z. This results in a 0.1% share for (γ,n)/(n,2n) for zirconium-90, see Table 11.

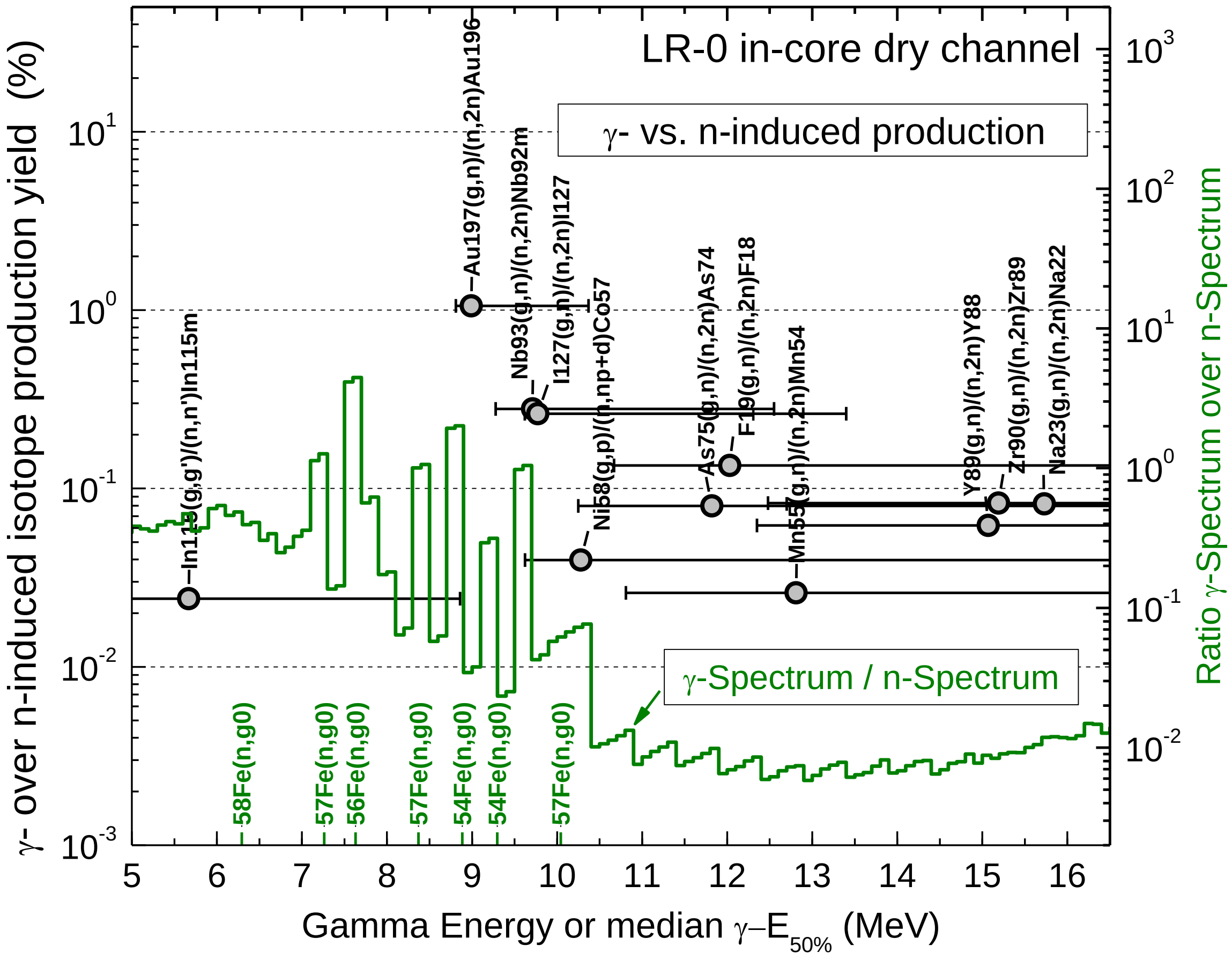

Figure 10.: Left axis: Calculated gamma induced over the neutron induced production yields for the same residual isotope versus the gamma median energy γ-$E_{50\%}$ (shown by points); the horizontal bars indicate the (5 – 95)% response domain. Right axis: Ratio of the absolute γ-ray and neutron spectra in the dry hexagonal channel of LR-0 - green histogram. The energies of γ-rays $E\gamma_0$ from ($n_{th}$,$\gamma_0$) on the Fe stable isotopes are indicated by green bars and labels

Table 11: Neutron and gamma induced reactions leading to the same residual nucleus

| Reaction | $E_{thr,}$ MeV | $E_{50\%}$, MeV | SACS, mb | Rel. Unc., % | Reaction Rate, 1/kcode | Fraction, % |
|---|---|---|---|---|---|---|
| | | | Competition between reactions (n,n') and (γ,γ') | | | |
| $^{115}$In(n,n')$^{115m}$In | 0.339 | 2.481 | 5.594E+01 | 1.69 | 1.688E-05 | 99.98 ± 0.00 |
| $^{115}$In(γ,γ')$^{115m}$In | 0.336 | 5.669 | 2.651E-02 | 0.03 | 4.065E-09 | 0.02 ± 0.00 |
| | | | Competition between reactions (n,2n) and (γ,n) | | | |
| $^{197}$Au(n,2n)$^{196}$Au | 8.114 | 10.420 | 9.113E-01 | 1.93 | 2.749E-07 | 98.96 ± 0.02 |
| $^{197}$Au(γ,n)$^{196}$Au | 8.072 | 8.992 | 1.892E-02 | 0.10 | 2.902E-09 | 1.04 ± 0.02 |
| $^{93}$Nb(n,2n)$^{92m}$Nb | 9.063 | 11.210 | 1.249E-01 | 0.88 | 3.767E-08 | 99.72 ± 0.00 |
| $^{93}$Nb(γ,n)$^{92m}$Nb | 8.966 | 9.713 | 6.861E-04 | 0.17 | 1.052E-10 | 0.28 ± 0.00 |
| $^{127}$I(n,2n)$^{126}$I | 9.217 | 11.460 | 3.279E-01 | 3.16 | 9.892E-08 | 99.74 ± 0.01 |
| $^{127}$I(γ,n)$^{126}$I | 9.144 | 9.773 | 1.690E-03 | 0.19 | 2.592E-10 | 0.26 ± 0.01 |
| $^{55}$Mn(n,2n)$^{54}$Mn | 10.414 | 12.810 | 6.893E-02 | 2.48 | 2.080E-08 | 99.97 ± 0.00 |
| $^{55}$Mn(γ,n)$^{54}$Mn | 10.227 | 12.810 | 3.523E-05 | 0.57 | 5.403E-12 | 0.03 ± 0.00 |
| $^{75}$As(n,2n)$^{74}$As | 10.383 | 12.820 | 9.031E-02 | 6.03 | 2.725E-08 | 99.92 ± 0.00 |
| $^{75}$As(γ,2n)$^{74}$As | 10.245 | 11.820 | 1.419E-04 | 0.42 | 2.176E-11 | 0.08 ± 0.00 |
| $^{89}$Y(n,2n)$^{88}$Y | 11.612 | 13.780 | 4.816E-02 | 1.37 | 1.453E-08 | 99.94 ± 0.00 |
| $^{89}$Yγ,n)$^{88}$Y | 11.482 | 15.070 | 5.883E-05 | 0.89 | 9.021E-12 | 0.06 ± 0.00 |
| $^{19}$F(n,2n)$^{18}$F | 10.986 | 13.880 | 2.252E-03 | 3.04 | 6.793E-10 | 99.87± 0.00 |
| $^{19}$F(γ,2n)$^{18}$F | 10.432 | 12.030 | 5.968E-06 | 0.54 | 9.151E-13 | 0.13 ± 0.00 |
| $^{90}$Zr(n,2n)$^{89}$Zr | 12.103 | 14.290 | 2.948E-02 | 1.08 | 8.895E-09 | 99.93 ± 0.00 |
| $^{90}$Zr(γ,n)$^{89}$Zr | 11.968 | 15.190 | 4.802E-05 | 0.91 | 7.364E-12 | 0.08 ± 0.00 |
| $^{23}$Na(n,2n)$^{22}$Na | 12.965 | 15.410 | 1.072E-03 | 1.54 | 3.235E-10 | 99.92 ± 0.00 |
| $^{23}$Na(γ,2n)$^{22}$Na | 12.216 | 15.730 | 1.727E-06 | 1.09 | 2.648E-13 | 0.08 ± 0.00 |
| | | | Competition between reactions (n,d+np) and (γ,p) | | | |
| $^{58}$Ni(n,np+d)$^{57}$Co | 6.051 | 12.663 | 3.281E-02 | 0.28 | 9.898E-09 | 99.93 ± 0.00 |
| $^{58}$Ni(γ,p)$^{57}$Co | 8.172 | 10.280 | 4.829E-05 | 0.28 | 7.404E-12 | 0.07 ± 0.00 |
| | | | Competition between reactions (n,α) and (γ,$^{3}$He) | | | |
| $^{63}$Cu(n,α)$^{60}$Co | 0.000 | 7.008 | 1.416E-01 | 3.07 | 4.271E-08 | 100.00 ± 0.00 |
| $^{63}$Cu(γ,$^{3}$He)$^{60}$Co | 18.861 | 21.250 | 1.812E-20 | 99.92 | 2.779E-27 | 0.00 ± 0.00 |
| $^{54}$Fe(n,α)$^{51}$Cr | 0.000 | 6.909 | 2.560E-01 | 3.70 | 7.722E-08 | 100.00 ± 0.00 |
| $^{54}$Fe(γ,$^{3}$He)$^{51}$Cr | 19.734 | 21.250 | 9.211E-21 | 100.00 | 1.412E-27 | 0.00 ± 0.00 |
| $^{27}$Al(n,α)$^{24}$Na | 3.249 | 8.482 | 1.889E-01 | 0.74 | 5.700E-08 | 100.00 ± 0.00 |
| $^{27}$Al(γ,$^{3}$He)$^{24}$Na | 23.710 | 5.523 | 1.232E-08 | 0.03 | 1.889E-15 | 0.00 ± 0.00 |
| $^{51}$V(n,α)$^{48}$Sc | 2.095 | 10.420 | 9.113E-01 | 1.93 | 2.749E-07 | 100.00 ± 0.00 |
| $^{51}$V(γ,$^{3}$He)$^{48}$Sc | 22.631 | - | - | - | - | - |

Competition between the neutron and gamma induced reactions leads to the same residual nucleus for the case of irradiation in the LR-0 in-core dry hexagonal channel. Columns contain neutron and γ-ray kinematic threshold $E_{thr,}$ median response energies $E_{50\%}$, calculated SACS with total relative uncertainties, production reaction rates normalized per 1 kcode event and production fraction for the reaction residual nucleus.

# 5 Conclusions

The SACS of the $^{14}$N(n,p)$^{14}$C reaction was measured in the LR-0 reactor benchmark reference neutron field. Calculations performed with the current major evaluated cross section libraries are in good agreement with the result of present experiments. This is very important for the quantitative assessment of the origin of $^{14}C$ from nitrogen. It is recommended to verify this single existing experiment by employing alternative methods, such as mass spectrometry.

The measured spectrum averaged cross sections of 30 reactions were compared with the calculated SACSs using different codes and cross section libraries, including IRDFF-II dedicated to dosimetry purposes. It was found that the agreement is usually within the uncertainties resulting from the measurements and those in the evaluated cross section libraries. The established set of dosimetry reactions can be efficiently used for testing the energy spectrum deconvolution codes.

The precise Monte-Carlo simulations and analytical estimation demonstrate the equivalence of the $^{235}U(n_{th},f)$ PFNS and the LR-0 spectrum above 6 MeV. Local oscillating deviations were shown to be caused by the quasi-resonances in the oxygen neutron total cross section. Selected LR-0 spectrum averaged cross sections were corrected to SACS averaged in $^{235}U(n_{th},f)$ PFNS. The resulted values are in very good agreement with existing values measured in other laboratories.

The new calculation of gamma spectrum up to 20 MeV is supported through validation with the measured gamma spectrum in 1–13 MeV region. Good agreement of measured values with calculation is observed, thus the calculations are suitable in evaluation of gamma. Competition between the gamma- and neutron-induced reactions leading to the same residual nucleus was calculated for most of benchmarked reactions. The maximum impact of ≈ 1% was observed for (γ,n)/(n,2n) in case of $^{196}Au$ and ≤ 0.3% for $^{93}Nb$ and $^{127}I$. For other competing reaction pairs and targets nuclei, the impact is smaller by one or more orders of magnitude.

The SACS for the $^{58}Ni(n,x)^{57}Co$ reaction cross section was also measured. The observed value is in quite good agreement with our previous measurement in another reactor (VR-1). This confirms the reliability of the experimental results and supports the recommendation to use this reaction as a new dosimeter.

## Acknowledgements

The presented results were obtained using the CICRR infrastructure, which is financially supported by the Ministry of Education, Youth and Sports - project LM2023041, the SANDA project funded under H2020-EURATOM-1.1 contract 847552 and by the Technology Agency of the Czech Republic within the National Competence Centre Program, project TN02000012 „Center of Advanced Nuclear Technology II“ which is partially co-financed within the National Recovery Plan from the European Instrument for recovery and resilience.